\documentclass[aip,pof,reprint,amsmath,amssymb,floatfix]{revtex4-1}

\usepackage{graphicx}
\usepackage{amsfonts,amssymb,amsmath,bm}

\providecommand{\gtrsim}{\:\raisebox{.25ex}{$>$}\hspace*{-.75em}
\raisebox{-.93ex}{$\sim$}\:}
\providecommand{\lesssim}{\:\raisebox{.25ex}{$<$}\hspace*{-.75em}
\raisebox{-.93ex}{$\sim$}\:}

\newlength{\myfigwidth}
\begin{document}

\title{On the theory and numerical modeling of spall fracture in pure liquids}

\author{Mikhail M. Basko}
\email{mmbasko@gmail.com}
\homepage{http://www.basko.net}
\affiliation{Keldysh Institute of Applied Mathematics, Miusskaya square 4, 125047 Moscow, Russia}

\date{\today}

\begin{abstract}
A novel phase-flip model is proposed for thermodynamically consistent and computationally efficient description of spallation and cavitation in pure liquids within the framework of ideal hydrodynamics. Aiming at ultra-fast dynamic loads, the spall failure of a liquid under tension is approximated as an instantaneous decomposition of metastable states upon reaching the spinodal stability limit of an appropriate two-phase liquid-gas equation of state. The spall energy dissipation occurs as entropy jumps in two types of discontinuous solutions, namely, in hypersonic spall fronts and in pull-back compression shocks. Practical application of the proposed model is illustrated with numerical simulations and a detailed analysis of a particular problem of symmetric plate impact. The numerical results are found to be in good agreement with the previously published molecular-dynamics simulations. Also, new approximate nonlinear formulae are derived for evaluation of the strain rate, fractured mass, and spall strength in terms of the observed variation of the free-surface velocity. The new formula for the spall strength clarifies complex interplay of the three first-order nonlinear correction terms and establishes a universal value of the correction factor for attenuation of the spall pulse, which in the limit of weak initial loads is independent of the equation of state.


\end{abstract}

\maketitle

\section{Introduction \label{s:intr}}

For many decades, investigation of spall failure in matter under dynamic tension was conducted mainly for solids \cite[]{Antoun_Seaman.2003, Kanel_Fortov.2007}. In recent years, measurements of the spall strength in liquid metals, loaded by the traditional method of impactor plates, have been reported \cite[]{Kanel_Savinykh.2015, Zaretsky2016}.  Also, many new experiments have been performed where spall failure occurs in liquids, often in molten metals, at very high loading rates under the action of nanosecond and sub-picosecond laser pulses \cite[]{Resseguier_Signor.2007, Agranat_Anisimov.2010, Krivokorytov_Vinokhodov.2017}, or very short X-ray bursts \cite[]{Stan_Willmott.2016}. Adequate modeling of dynamic spall fracture in liquids is a challenging problem, especially when the geometry of the experiment is not one-dimensional \cite[]{Stan_Willmott.2016} and/or spall fracture is only part of a more complex problem as, for example, by modeling liquid tin targets for EUV lithography sources \cite[]{Basko_Krivokor.2017, Grigoryev_Lakatosh.2018}.

A well-established approach is to use the first-principle molecular dynamics (MD) simulations \cite[]{Pisarev_Kuksin.2009,Cai_Wu.2017, Mayer_Mayer2020}, which provide an important insight and allow calculation of the ultimate spall strength in liquids at high strain rates but are limited to sub-micron targets, are computationally very costly and, for that reason, not quite practical. A much simpler and computationally inexpensive alternative is to simply adopt the fully equilibrium (EQ) equation of state (EOS), obtained by applying Maxwell's rule to the primitive EOS of the van-der-Waals (vdW) type \cite[]{Anisimov_Inogamov.1999, Colombier_Combis.2005, Zhao_Mentrelli.2011, Basko_Krivokor.2017}. However, because the EQ EOS allows no negative pressures, it cannot describe tensile states, pull-back shocks, and only approximately reproduces mass distribution in the regions of spallation and cavitation \cite[]{Basko2018-PhF}. Another often used model \cite[]{Boteler_Sutherland2004, Grigoryev_Lakatosh.2018} is based on combining an appropriate empirical EOS that admits negative pressures with an ad hoc limit on the tensile stresses: once and where the spall threshold is exceeded, an expanding void cavity with zero-pressure boundaries is introduced into the simulation. The principal flaw of this approach is its inability to adequately account for the liquid-vapor phase transition, especially when spall failure occurs practically simultaneously over a large quasi-uniform mass of metastable liquid, converting it into a finely dispersed liquid-vapor mixture.

In the present work, we investigate a novel phase-flip (PF) model based on the premise that all the information, needed for modeling cavitation and spallation in pure liquids, is provided by a two-phase liquid-gas vdW-type EOS, and no other material properties such as surface tension, viscosity, etc.\ are to be invoked. This is a clear aspect where our model differs from the energy-based scheme by Grady\cite{Grady1988, Grady1996}, preserving nonetheless the capability to adequately account for the energy dissipation by spallation. Aiming at description of ultra-fast dynamic processes, we assume the spall strength to attain its maximum possible value, i.e.\  to be defined by the thermodynamic stability limit of the vdW-type EOS along the spinodal curve. The spall fracture is then naturally associated with the spinodal decomposition, which, in line with the adopted minimalistic approach, is assumed to take place instantly. To justify the latter assumption, we invoke the theory of homogeneous bubble nucleation in metastable liquids.

The primary motivation for proffering the PF model is its potential to be easily and computationally cheaply integrated into large multi-dimensional fluid dynamics codes that are already loaded with many other complex physical processes like spectral radiation transfer, laser energy deposition, etc. when need arises to describe target fragmentation by spallation and cavitation under fast and intense initial loads. A notable example is the task to adequately describe the initial shattering of small liquid-tin drops by picosecond laser pulses in EUV-lithography applications \cite[]{Bakshi2006, Mizoguchi_Nakari.2015}, where MD codes cannot be directly applied, and where the PF model promises to become a major improvement over the more primitive models employed in Refs.~\onlinecite{Basko_Krivokor.2017, Grigoryev_Lakatosh.2018}.

The full formulation and justification of the PF model is given in Sect.~\ref{s:PFM}, where we start by introducing the employed vdW-type EOS, describe and analyze the adopted spall criterion, and explore the kinematics of hypersonic spall fronts. To illustrate how the proposed model could be applied in practice, we perform a detailed analysis of the spallation process in a planar collision between identical flyer and target plates. Although real experiments are usually done for asymmetric configurations with target-to-flyer thickness ratios about 3--5 \cite[]{Antoun_Seaman.2003}, we choose the theoretically simplest symmetric case to display the key properties of our model, and to compare it with the available MD simulations.
Our numerical results, presented in Sect.~\ref{s:num}, indicate that the PF model can reproduce the key aspects of the spall dynamics no less accurately than the MD simulations --- provided that an adequate two-phase EOS is available. The difference emerges only in the microstructure of fracture zones, stipulated by statistical atomic-scale properties of an MD model, and determined by small-scale perturbations in the PF model.

Interpretation of plate impact experiments for determination of the spall strength $\sigma_{sp}$ relies on the theoretical relationship between $\sigma_{sp}$ and the observed variation $\Delta u_{fs}$ of the free-surface velocity  (the ``velocity pullback''), for which usually the linear acoustic approximation is used. In Sect.~\ref{s:PA} we develop a new variant of the post-acoustic approximation and derive a new formula relating $\sigma_{sp}$ to $\Delta u_{fs}$ in the case of a symmetric plate impact. The new formula is significantly more accurate than the acoustic one, and provides an important insight into the complex interplay of the first-order nonlinear correction terms. New post-acoustic formulae are also obtained for the strain rate and the fractured mass fraction.

\section{The phase-flip model of spall fracture in liquids \label{s:PFM}}

\subsection{The employed two-phase EOS \label{s:eos}}

Conceptually, the thermodynamics of a liquid-gas phase transition is adequately represented by the classical van der Waals EOS. In this work we make use of one of the simplest generalizations of this EOS (termed GWEOS) \cite[]{Martynyuk1991, Martynyuk1993, Basko2018} in the form
\begin{eqnarray}\label{p(v,tet)=}
  p(v,\theta) &=& \frac{\alpha \theta}{v-\kappa^{-1}} -\frac{\kappa}{v^n},
  \\ \label{e(v,tet)=}
  e(v,\theta) &=& \alpha c_V \theta -\frac{1}{2} \kappa(\kappa-1) v^{1-n}, \\ \label{s(v,tet)=}
  s(v,\theta) &=& \alpha\left[c_V(1+\ln\theta) +\ln(v-\kappa^{-1}) \right],
\end{eqnarray}
where $\kappa=(n+1)/(n-1)$, $\alpha= \kappa-\kappa^{-1}$, and the exponent $n>1$ together with a constant heat capacity $c_V>0$ are its two free dimensionless parameters. The EOS (\ref{p(v,tet)=})--(\ref{s(v,tet)=}) is cast in its reduced form, where the pressure $p$, the specific volume $v \equiv \rho^{-1}$ and the temperature $\theta$ are normalized to the corresponding values $P_{cr}$, $V_{cr}= \rho_{cr}^{-1}$ and $T_{cr}$ at the liquid-gas critical point; the mass-specific internal energy $e$ is in units of $P_{cr}V_{cr}$, the entropy $s$ in units of $P_{cr}V_{cr}/T_{cr}$. The dimensional quantities $P_{cr}$, $V_{cr}$, $T_{cr}$ make up additional three free parameters of GWEOS; for more details see Ref.~\onlinecite{Basko2018}. The main motivation for using GWEOS in this work is its mathematical simplicity, which allows the in-line use of its both the metastable (MS) and the equilibrium (EQ) branches in hydrodynamic codes with the rounding-error accuracy --- thus excluding numerical errors due to imperfect EOS.

In numerical examples, discussed below, the few free parameters of GWEOS were chosen such as to provide the best fit to the properties of water near normal conditions. The well known critical parameters of water allow confident normalization of its other measured quantities to the system of reduced variables. For simplicity, we assume the normal state $(v_0,\theta_0)$ to have zero pressure $p(v_0,\theta_0)=0$. The values of $n$, $c_V$ and $\theta_0$ are adjusted to fit as best as possible the experimental values of the normal specific volume $v_0$, the isentropic sound speed $c_0$, the vaporization enthalpy $h_{vap}$, and the specific heat $\alpha c_V$ at $T=20^{\circ}$~C. Table~\ref{t:1} shows how good a compromise has been achieved with the values
\begin{equation}\label{n=}
  n=1.5, \quad c_V=8.18, \quad \theta_0=0.45,
\end{equation}
used throughout this paper.

\begin{table}[hbt]
\caption{Comparison between the normal properties of water  \cite[]{CRC2015} and those produced by GWEOS with parameters~(\ref{n=}). For all quantities the reduced values are given}
\label{t:1}
\begin{ruledtabular}
\begin{tabular}{lllllll}
  & $\theta_0$ & $\rho_0$ & $c_0$ & $h_{vap}$ & $\alpha c_V$ & $b_0$ \\
  \hline
   GWEOS & 0.450 & 3.9073 & 5.9939 & 21.5 & 39.26 & 3.17 \\
   H${}_2$O & 0.453 & 3.10 & 5.66 & 35.8 & 39.26 & 2.0 \\
\end{tabular}
\end{ruledtabular}
\end{table}

\begin{figure}[hbt]
  \centering
  \includegraphics[width=\myfigwidth]{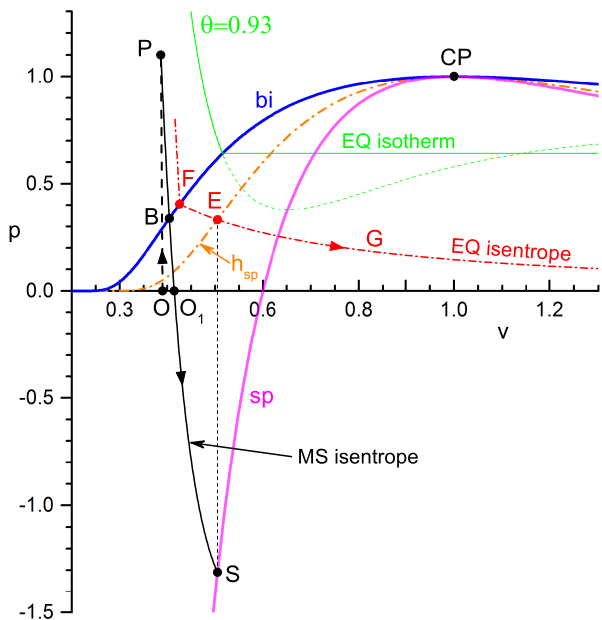}
  \caption{Thermodynamic $(v,p)$ plane of a fluid with a liquid-gas phase transition. Shown are the binodal $bi$ and the spinodal $sp$ curves with the critical point $CP$ at $(v,p)=(1,1)$. Curve $PBO_1S$ shows an MS isentrope, along which the release-wave states would evolve after being brought to $P$ along the Hugoniot $OP$ (dashed) starting from the normal state $O$.  Dash-dotted curve $h_{sp}$ (orange) is the locus of isochoric post-flip states $E$ as the initial state $S$ slides along the spinodal $sp$; dash-dotted curve $FEG$ (red) is the EQ isentrope passing through the chosen state $E$. The behavior of the MS (solid green) and EQ (dashed green) isotherms is illustrated for $\theta=0.93$} \label{f:1}
\end{figure}

As is shown in Fig.~\ref{f:1}, Eq.~(\ref{p(v,tet)=}) exhibits all the standard features of the vdW EOS. The phase coexistence region lies between the \emph{binodal} curve $bi$ and the $p=0$ line. The EQ EOS in this region, denoted $p_{EQ}(v,\theta)$, $e_{EQ}(v,\theta)$, $s_{EQ}(v,\theta)$, and the position of the binodal itself are determined by applying the Maxwell rule \cite[\S\S 84,85]{LL-SP96} to the MS isotherms (\ref{p(v,tet)=}), as is illustrated with the horizontal segment of the $\theta=0.93$ isotherm in Fig.~\ref{f:1}. The \emph{spinodal} curve $sp$, defined by the condition $\partial p(v,\theta)/\partial v =0$, lies under the binodal and delimits the region of absolute thermodynamic instability; hence, only the EQ states with $p>0$, obtained by applying the Maxwell rule, are possible below the spinodal. In the intermediate metastable region between the binodal and the spinodal both the MS and the EQ branches of EOS are thermodynamically admissible and compatible with the equations of fluid dynamics. The latter is ensured by the positiveness of the square of the isentropic sound speed
\begin{equation}\label{c^2=}
  c^2 = \left(\frac{\partial p}{\partial\rho} \right)_s >0
\end{equation}
everywhere along and above the spinodal curve. Tension states with $p<0$ are only possible for the MS states in the superheated-liquid region at $v<1$. Note that, by assuming $p(v_0,\theta_0)=0$, we put the normal state $(v_0,\theta_0)$ into the category of metastable ones.

A non-trivial point is the interpretation of the EQ states from the viewpoint of matter structure. While there is no ambiguity about the metastable liquids, which can always be assumed homogeneous, the EQ states are normally believed to be necessarily heterogenous \cite[\S 83]{LL-SP96}. On the other hand, no length scale for their heterogeneity can be derived from our basic thermodynamic relations (\ref{p(v,tet)=}), (\ref{e(v,tet)=}). Without extra elements of physics (like the kinetics of bubble nucleation, surface tension, impurities, etc.), one can only assume that an EQ state represents a quasi-uniform finely dispersed liquid-gas mixture, where the characteristic sizes of either vapor bubbles or liquid drops are always much smaller than the smallest relevant length scale in the considered problem [see, for example, problem~1 to \S~64 in~\cite[]{LL-H87}]. Below we apply the term ``fog'' to  such EQ states and treat them as perfectly homogeneous in the context of spallation and cavitation in pure liquids.

\subsection{Spall criterion and the phase-flip approximation \label{s:Gr}}

Spall failure inside (i.e.\ away from boundaries) a volume of pure liquid occurs in the process of spontaneous decay of its metastable state due to thermodynamic fluctuations \cite[]{Grady1988}. This implies that, other factors being equal, the spall strength $\sigma_{sp}$ must increase with the increasing strain rate, which is confirmed by many experiments \cite[]{Kanel_Fortov.2007, Kanel2010} and the MD simulations \cite[]{Mayer_Mayer2020}. Hence, the theoretical maximum of $\sigma_{sp}$ --- the \emph{ultimate strength} --- must be attained at the spinodal
\begin{equation}\label{Gr:sig_sp=}
  \sigma_{sp} = -p_{sp}(v) =\frac{n-v(n+1)}{v^{n+1}}, \quad \frac{n-1}{n+1}< v <\frac{n}{n+1},
\end{equation}
where the metastability decay rate is maximum \cite[]{Imre_Drozd-Rzoska.2008}.

In this work we adopt a simple and universal criterion that a given fluid element undergoes spall failure whenever its thermodynamic trajectory reaches the spinodal from the side of metastable liquid. The results obtained under this assumption should be in the first place applicable to tensile loads with the highest possible strain rates, like, for example, $\dot{\varepsilon} \simeq 10^9$--$10^{10}$~s${}^{-1}$ observed in laser experiments. In reality, of course, the tensile strength can only approach the spinodal limit from below \cite[]{Imre_Drozd-Rzoska.2008}, but how closely --- about 0.3--0.5 \cite[]{Kanel_Fortov.2007, Agranat_Anisimov.2010} or up to 0.8--0.9 \cite[]{Pisarev_Kuksin.2009, Malyshev_Marin.2015} of the limiting value --- is still debatable. Given this uncertainty and the uncertainty in the spinodal position for real materials, we do not try to soften this criterion or make it dependent on $\dot{\varepsilon}$, the more so that it would not affect the main results of this work.

Having set the pressure threshold for spallation, we need to add further assumptions about the kinetics and morphology of the ensuing fracture. Because the rate of spontaneous bubble nucleation very rapidly increases by many orders of magnitude as the MS liquid approaches the spinodal \cite[]{Blander_Katz1975, Skripov_Skripov1979}, we can make a simplifying assumption that spall failure occurs as an instantaneous and irreversible MS~$\rightarrow$~EQ phase jump --- the \emph{phase flip}. The extremely short timescale, on the order of a few picoseconds, of spinodal decomposition in real liquids justifies this approximation for strain rates up to $\dot{\varepsilon} \simeq 10^{11}$~s${}^{-1}$. Because the rate of density variation $\dot{\varepsilon}$ is controlled by hydrodynamics (i.e.\ by fluid inertia and pressure gradients), the phase flip must occur at a constant $\rho=v^{-1}$ whenever $\dot{\varepsilon}$ is finite. The only possibility where it could be accompanied by a density jump, would be inside a rarefaction shock \cite[]{Basko2018-PhF}, which can be excluded in the context of spallation because relaxation of tension means growth of pressure. In addition, energy conservation requires the local value of the specific internal energy $e$ to be preserved as well, which implies that the phase flip is always accompanied by a jump-like increase in pressure, temperature, and entropy.

In Fig.~\ref{f:1} the trajectory $PBO_1SE$ illustrates possible evolution of the thermodynamic state of a fluid element, undergoing spall failure: expansion along the MS isentrope $PBO_1S$ brings it to a negative-pressure state $S$ on the spinodal, followed by an instantaneous jump to the EQ ``fog'' state $E$ at a higher entropy; its subsequent evolution is governed by the EQ EOS branch so long as it remains under the binodal. If the tracked mass of ``fog'' emerges from under the binodal (say, by recompression along the EQ isentrope $EF$), it may later return to the two-phase region again as metastable liquid and undergo a secondary spall failure. Thus, the fluid elements in our scheme can exhibit a hysteresis-like thermodynamic behavior that is fully consistent with the first and second laws of thermodynamics.

Because metastable liquid can border vacuum at a finite density, one might conjecture that it would rupture by forming an expanding vacuum cavity, originating from the \emph{inception point} where the spall threshold is first attained. To assess the feasibility of such scenario, we have to consider the compression shock pulse that would normally be launched into the surrounding stretched liquid by pressure relaxation in the fractured material (often called the ``spall pulse''). Because this shock relaxes the tension in metastable liquid and pulls its thermodynamic state back from the spall threshold, we call it the \emph{pull-back (pb) shock}. For an expanding vacuum cavity to form, the pb shock would have to start immediately from the inception point. However, a closer examination in the next section reveals a different general pattern: once the spall threshold is reached at the inception point, the region of spall failure very rapidly (hypersonically) extends over some finite volume, bounded by a surface from where the pb shock does actually start, and which may be called the ``spall horizon'' of a given inception point. Due to the continued  stretching, sustained by inertia of the fluid motion, no fluid element inside the spall horizon can be reached by the pb shock before it undergoes spall failure. In other words, the spall fracture in our model develops as a rapid phase transition between the MS and EQ EOS branches, occurring practically simultaneously throughout a certain finite mass. Since an EQ state can border vacuum only at zero density, there is no need to postulate any void openings, and the spalling (cavitating) liquid can be treated as everywhere continuous.

At first glance, the above spall criterion appears similar to that proposed earlier by Grady\cite{Grady1988, Grady1996}, who also assumed an instantaneous MS~$\rightarrow$~EQ phase jump in the framework of a vdW-like EOS. The  essential point of difference is, however, that Grady postulated the post-jump EQ state $E$ to have the same entropy as the pre-jump MS state $S$. As the EQ isentrope, branching from the same point $B$ on the binodal, has always lower values of the internal energy $e_{EQ}(v,s) < e(v,s)$ than the MS isentrope at the same density, Grady interpreted the excess energy $\Delta e_{NE} =e(v,s)- e_{EQ}(v,s)$ as the interfacial energy (extraneous to EOS) of the heterogeneous EQ state, consisting of liquid fragments with a characteristic size $l_{fr}$; he then used $\Delta e_{NE}$ and $l_{fr}$ to evaluate the spall strength by invoking the surface tension and viscosity. Our model, in contrast, is based on the assumption that $l_{fr}$ is very small and cannot be resolved hydrodynamically. Then the eventual interfacial energy must be automatically included into the fluid EOS, which requires the MS~$\rightarrow$~EQ phase jump to be iso-energetic and accompanied by the entropy increase.

Another highly controversial point in Grady's argumentation is the postulated linear dependence
\begin{equation}\label{Gr:l_fr=}
  l_{fr} =2ct_w
\end{equation}
of the typical fragment size $l_{fr}$ on the ``waiting'' time $t_w \propto \dot{\varepsilon}^{-1}$ elapsed  after crossing the binodal. Here we argue that such a linear relation would be in stark disagreement with the theory of homogeneous nucleation \cite[]{Skripov1974, Skripov_Skripov1979}, which predicts that the new phase in metastable liquids under tension $-p>0$ appears in the form of vapor bubbles with the critical radius
\begin{equation}\label{Gr:r_cr=}
  r_{cr} = \frac{2\alpha_{st}}{p_{sat}-p} \approx -\frac{2\alpha_{st}}{p},
\end{equation}
where $\alpha_{st}$ is the surface tension, and $p_{sat}= p_{sat}(T)>0$ is the saturated vapor pressure at the same temperature $T$ for which $p$ is calculated \cite[]{LL-SP96, Skripov1974, Blander_Katz1975}. Because the probability of spontaneous creation of such bubbles, proportional to $\exp(-G)$ where
\begin{equation}\label{Gr:G=}
   G= \frac{16\pi\alpha_{st}^3}{3(p_{sat}-p)^2T} = \frac{4\pi\alpha_{st} r_{cr}^2}{3T} \gg 1
\end{equation}
is the Gibbs number, is a very steep function of $p$ and $T$, the spall fracture of a uniformly stretched liquid occurs as a sudden emergence of myriad critical vapor bubbles with radii of a few nanometers \cite[]{Skripov_Skripov1979, Pisarev_Kuksin.2009}, which practically instantaneously (i.e.\ with virtually no time for subsequent growth) begin to coalesce, breaking the molecular bonds that sustain tension. Within this picture, it is expected that $l_{fr} \approx r_{cr}$ and $t_w \propto \exp(G)$, which, together with Eq.~(\ref{Gr:G=}), implies a highly nonlinear relation between $l_{fr}$ and $t_w \propto \dot{\varepsilon}^{-1}$, and significantly smaller values of $l_{fr}$ than predicted by Eq.~(\ref{Gr:l_fr=}). Therefore, when the goal is to explore the dependence of $\sigma_{sp}$ on the strain rate $\dot{\varepsilon}$, the simple relationship (\ref{Gr:l_fr=}) must be replaced by a more adequate model based on the nucleation theory --- as, for example, was proposed in~\cite[]{Dekel_Eliezer.1998, Pisarev_Kuksin.2009, Faik_Basko.2012}.

In summary, unlike Grady's scheme, our model describes energy dissipation due to spall fracture  without involving the surface tension and viscosity, simply as the entropy increase in the phase-flip transitions and behind the pull-back shock fronts. In this respect it is similar to the viscous-free ideal hydrodynamics with dissipation due only to discontinuous shocks. The weak dependence of the spall strength on the strain rate is ignored by adopting the theoretically maximum value of $\sigma_{sp}$ allowed by thermodynamics.

\subsection{Kinematics of spall fronts and the size of fracture zone \label{s:spf}}

Consider a flow of liquid with a velocity field $\mathbf{u}(t,\mathbf{x})$, where a certain region is stretched to develop negative pressures. The local rate of stretching is characterized by the \emph{strain rate}, defined as
\begin{equation}\label{spf:deps=}
  \dot{\varepsilon} \, \stackrel{\mbox{\scriptsize def}}{=} \, -\frac{d\ln\rho}{dt} = \frac{d\ln v}{dt} = \nabla\cdot \mathbf{u},
\end{equation}
where $d/dt = \partial/\partial t+ \mathbf{u}\cdot \nabla$ is the material time derivative, and the use is made of the fluid continuity equation
\begin{equation}\label{spf:drho/dt=}
  \frac{\partial\rho}{\partial t} + \nabla\cdot \left(\rho\mathbf{u}\right) =0.
\end{equation}

Next, we assume for simplicity that the flow is isentropic with a fixed value of entropy $s$ throughout the considered volume. Then the pressure becomes a known monotonic function $p(v) =p(v,s)$ of a single variable $v$ with $dp/dv<0$ everywhere above the spinodal (curve $PBO_1S$ in Fig.~\ref{f:1}); the spall strength $\sigma_{sp} = -p(v_{sp})$ and the spall density $\rho_{sp} =v_{sp}^{-1}$ take on universal values that are determined by the intersection of the isentrope $p(v)$ with the liquid branch of the spinodal (point $S$ in Fig.~\ref{f:1}).

\begin{figure}[hbt]
  \centering
  \includegraphics[width=\myfigwidth]{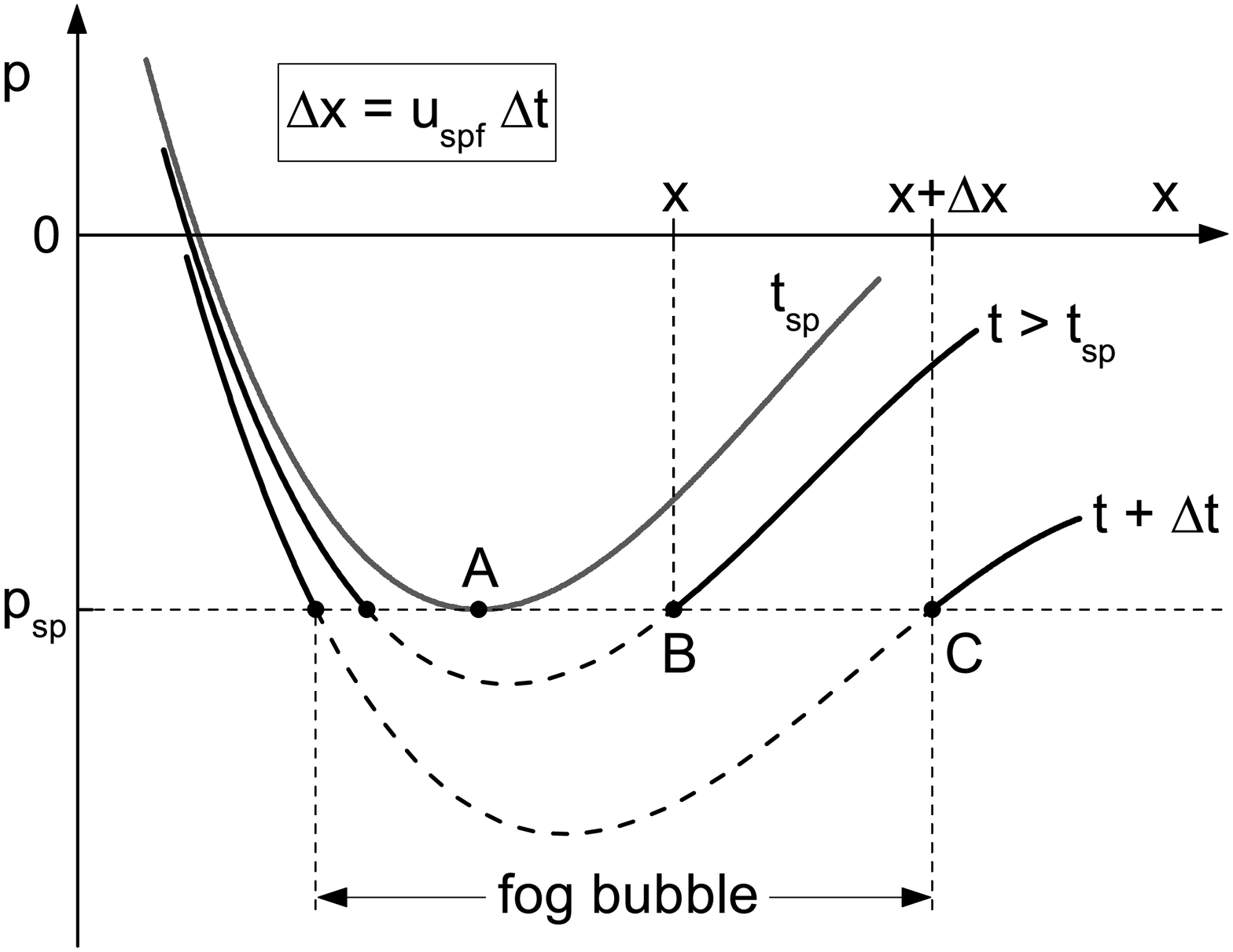}
  \caption{Schematics of the spall front propagation. As the local pressure minimum is forced to drop ever deeper below the spall threshold $p_{sp}$, the position of the spall front shifts from the inception point $A$ at $t_{sp}$ to points $B$ and $C$ at later times $t$ and $t+\Delta t$} \label{f:2}
\end{figure}

Spall failure is initiated at time $t_{sp}$ when the absolute minimum in the spatial pressure distribution $p(t_{sp},\mathbf{x})$ attains the threshold value $p_{sp}= -\sigma_{sp}$ at some inception point $A$, as is illustrated in Fig.~\ref{f:2}. Generally, when $p(t,\mathbf{x})$ and $p(v)$ are differentiable at $A$, one has $\nabla p=\nabla\rho =0$ at this point. At later times $t> t_{sp}$ the condition $p(t,\mathbf{x}) = p_{sp}$ will be satisfied on a certain surface around the inception point, which we call the \emph{spall front} and which encompasses a fog bubble. As suggested by the term, all the liquid inside the fog bubble has been, upon passing through the spall front, instantaneously transformed into an EQ ``fog'' state at a positive pressure. The condition $\nabla p=0$ by inception implies that the expansion of the fog bubble starts infinitely fast, but gradually slows down as $\left|\nabla p\right|$ increases. Because $p$ is a single-valued monotonically growing function of $\rho$, the normal (i.e.\ along the colinear $\nabla p$ and $\nabla\rho$ vectors) kinematic velocity of the spall front
\begin{equation}\label{spf:u_spf=}
  u_{spf} = \frac{\mathbf{u}\cdot\nabla\rho}{\left|\nabla\rho\right|} + \frac{\dot{\varepsilon}}{\left| \nabla\ln\rho\right|}
\end{equation}
is readily calculated by differentiating the equation $\rho(t,\mathbf{x}) = constant$, as is illustrated in Fig.~\ref{f:2}.  Note that in the comoving reference frame, where $\mathbf{u}=0$, the front velocity $u_{spf}$ is given by the second term on the right-hand side of Eq.~(\ref{spf:u_spf=}).

The merely kinematic propagation of the spall front with velocity (\ref{spf:u_spf=}) continues until it is overtaken by the pb shock, generated by the pressure jump $\Delta p_{pb} = p_{EQ,sp}+\sigma_{sp}$ behind the spall front; here $p_{EQ,sp}>0$ is the fog pressure in the post-flip EQ state (point $E$ in Fig.~\ref{f:1}), calculated from the EQ EOS branch for the same values of $v_{sp}$ and the internal energy $e_{sp}$ as in the pre-flip MS state $(v_{sp},p_{sp})$ (point $S$ in Fig.~\ref{f:1}). The pb-shock front is locally coplanar with the spall front and propagates in the same direction with the normal velocity
\begin{equation}\label{spf:u_pbf=}
  u_{pbf} = \frac{\mathbf{u}\cdot\nabla\rho}{\left|\nabla\rho\right|} + D_{pb},
\end{equation}
where $D_{pb} =D_{pb}(s)$ is the velocity of the shock front, driven by the pressure jump $\Delta p_{pb}$, relative to the liquid in the pre-shock state $(v_{sp},p_{sp})$; it is, of course, calculated by using the MS EOS branch. Similar to $v_{sp}$, $\sigma_{sp}$, and $\Delta p_{pb}$, the shock velocity $D_{pb}$ is uniquely determined by the entropy value $s$.

Once the condition $u_{pbf} \geq  u_{spf}$ is met and the pb shock breaks out before the spall front, it recompresses and heats up the stretched liquid, preventing it from reaching the spall threshold. As a result, the expansion of the fog bubble in Lagrangian coordinates, i.e.\ relative to the fluid particles, comes to a halt. In other words, the mass of fractured liquid continues to grow so long as the effective \emph{spall Mach number}, defined as
\begin{equation}\label{spf:Mach_sp=}
  \mathcal{M}_{sp}
  \, \stackrel{\mbox{\scriptsize def}}{=} \,
  \left( \frac{u_{spf}}{u_{pbf}} \right)_{\mathbf{u}=0} =
  \frac{\dot{\varepsilon}}{D_{pb} \left|\nabla \ln\rho \right|} = \frac{\dot{\varepsilon}\rho c^2}{D_{pb} \left|\nabla p\right|},
\end{equation}
stays above unity, $\mathcal{M}_{sp} > 1$. Note that $\mathcal{M}_{sp}$ is defined in the comoving frame, and relative not to the local sound speed $c$ but to the pb-shock velocity $D_{pb}>c$; hence, we use the term \emph{hypersonic} to a spall front with $\mathcal{M}_{sp} > 1$. The condition $\mathcal{M}_{sp}=1$, when fulfilled on the spall-front surface, delimits the size and mass of the fog bubble. The bubble mass is always finite when $\dot{\varepsilon}> 0$ and $\nabla\rho= \nabla p =0$ at the inception point.  It may become zero in some singular cases where either $\dot{\varepsilon}$ changes sign or $\nabla p$ is discontinuous at inception. In practice, a fair estimate of the fractured mass can often be made by analyzing the spatial distribution of $\mathcal{M}_{sp}(t_{sp},\mathbf{x})$ at the moment of inception $t_{sp}$. The above arguments and formulae can be readily generalized to non-isentropic flows, where $\mathcal{M}_{sp}$ becomes a function of gradients of two independent thermodynamic variables.

\section{Numerical results for a shock-loaded planar target \label{s:num}}

\subsection{Formulation of the problem and the initial state \label{s:ist}}

In a common experiment for determination of the spall strength, two planar plates are collided to launch the initial load shocks in two opposite directions from the impact plane \cite[]{Antoun_Seaman.2003}. In this work we focus our attention on the simplest version of such an experiment, where both plates have the same composition and dimensions. Then the problem is one-dimensional (1D) and symmetrical with respect to the impact plane, which we place at the origin $x=0$ of the center-of-mass coordinate system and consider only one (right) plate at $ x \geq 0$, which we refer to as the target. We set $t=0$ at the moment when the load shock breaks out at the outer edge $x=h_1$, leaving behind the shocked material in a compressed state at zero initial velocity. Thus defined initial state $(v_1,p_1)$ (point $P$ in Fig.~\ref{f:1}) lies on the Hugoniot $OP$, originating from the normal state $(v_0,p_0)$ at point $O$ in Fig.~\ref{f:1}. The intensity of the initial load is characterized by the particle (piston) velocity $u_p$, measured relative to the fluid in front of the loading shock. The latter means that, before the impact, the flyer and the target plates had, respectively, the velocities $+u_p$ and $-u_p$ in the center-of-mass frame; their density and thickness were $\rho_0$ and $h_0$. The value of $u_p$ uniquely determines the entropy $s$ throughout the subsequent flow, the spall strength $\sigma_{sp}$, as well as the pressure amplitude $\Delta p_{pb}$ and the front velocity $D_{pb}$ of the pb shock. Table~\ref{t:2} lists the key flow parameters that can be calculated directly from EOS for two chosen reference cases of the shock load.

\begin{table*}
\caption{Shock-load and subsequent flow parameters that can be calculated directly from EOS for two selected values of the post-shock particle velocity $u_p$. For detailed explanations see the text}
\label{t:2}
\begin{ruledtabular}
\begin{tabular}{llllllllllll}
 $u_p$ & $u_1$ & $\rho_1= v_1^{-1}$ & $c_1$ & $\rho_{01}$ & $c_{01}$ & $b$ & $z_0$ & $p_{EQ,sp}$ & $D_{pb,sp}$ & $c_{sp}$ & $\sigma_{sp}$ \\
  \hline
  1.25 & 1.2170 & 4.4470 & 14.351 & 3.8928 & 5.9329 & 3.1405 & 1.0140 & 0.00256 & 4.144 & 1.266 & 11.149 \\
  4.0  & 3.7495 & 4.7919 & 46.732 & 3.6309 & 4.9757 & 2.6954 & 1.2963 & 0.01302 & 3.635 & 1.241 & 8.2190 \\
\end{tabular}
\end{ruledtabular}
\end{table*}

Before specifying the initial target thickness $h_1$ (or its thickness $h_0$ before the shock load), we remark the following. One easily verifies that the above formulated problem is invariant with respect to rescaling of all the lengths and times by one and the same arbitrary factor. Therefore, having obtained the full solution for an initial target thickness $h_1$, we can rescale it to any other thickness $h_1'$ by multiplying times and lengths by the factor $h_1'/h_1$. In view of such scalability, all the simulations discussed below were done for $h_1=1$. Recalling that $v$, $p$, and $u$ are, respectively, everywhere in units of $V_{cr}$, $P_{cr}$, and $\left(P_{cr}V_{cr}\right)^{1/2}$, the latter is equivalent to setting the units of length and time to $h_1$ and $h_1\left(P_{cr}V_{cr}\right)^{-1/2}$.

\subsection{Simulation results for $u_p=1.25$ and $u_p=4.0$ \label{s:sim}}

\begin{figure}[hbt]
  \centering
  \includegraphics[width=\myfigwidth]{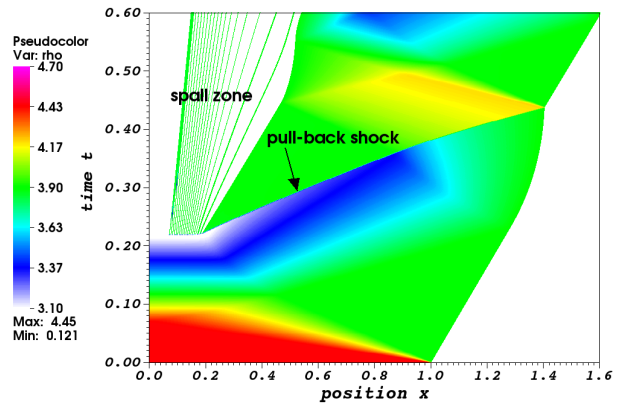}
  \caption{Color density map on the space-time diagram as calculated with the 1D DEIRA code for $u_p=1.25$} \label{f:3}
\end{figure}

Given an adequate two-phase EOS, implementation of our spall criterion into a standard Lagrangian hydrocode is straightforward. Here we present numerical results, obtained with the 1D DEIRA code that has been previously developed and extensively used for simulations of inertial confinement fusion targets \cite[]{Basko2001-DEIRA,Basko1990}. Figures~\ref{f:3} and \ref{f:5} display the target density evolution on the $(x,t)$ plane calculated, respectively, for $u_p=1.25$ and $u_p=4.0$. For a detailed discussion, we focus on the $u_p=1.25$ case, which lies moderately above the spall threshold of $u_{p,sp} \approx 0.82$. The high-load case of $u_p=4.0$ mainly serves to make the nonlinear effects more conspicuous

As the centered rarefaction wave, starting at $x=1$, arrives at and is reflected from the symmetry plane $x=0$, a tension region with $\rho < 3.90$ develops near the target center at $t \gtrsim 0.12$. Further on, as the pressure in this region falls to $p= -\sigma_{sp} =-11.15$, all the liquid mass at $0< x \lesssim 0.19$ undergoes spall fragmentation within a short time interval of $t_{sp}=0.21819< t< 0.220$, making up the spall zone that continues to expand preserving its mass. A salient feature is the pb shock, launched from the outer boundary of the spall zone at $t= 0.220$, and giving a kick to the free surface velocity later at $t=0.434$.

\begin{figure}[hbt]
  \centering
  \includegraphics[width=\myfigwidth]{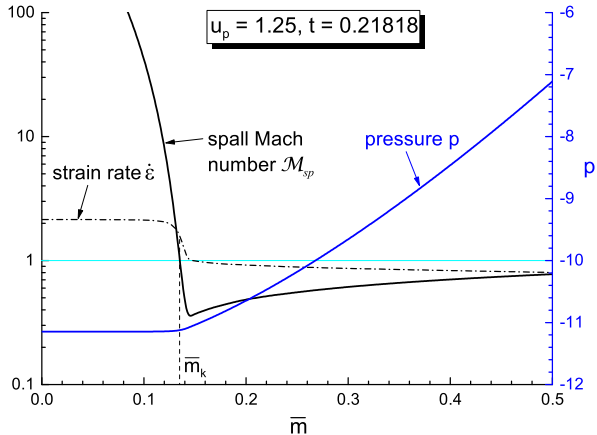}
  \caption{Profiles of the strain rate $\dot{\varepsilon}$, the spall Mach number $\mathcal{M}_{sp}$ (left ordinates), and of the pressure $p$ (right ordinates) versus the fractional target mass $\bar{m}$ just before the spall failure for $u_p=1.25$. Only the inner half of the target mass is shown} \label{f:4}
\end{figure}

More insight into the spall dynamics can be gained by examining the profiles in Figs.~\ref{f:4} and \ref{f:6} along the dimensionless Lagrangian coordinate
\begin{equation}\label{num:bar_m=}
  \bar{m} =\frac{m}{m_0}, \quad
   m=\! \int\limits_0^x \rho(t,x')\, dx', \quad
   m_0 = \rho_0h_0  = \rho_1h_1,
\end{equation}
plotted for the time moment immediately preceding $t_{sp}$. (Note that these profiles are invariant with respect to rescaling of the target thickness.) In Fig.~\ref{f:4} the spall failure occurs at $t_{sp}= 0.21819$ in the center $\bar{m}=0$, where the pressure is minimum and $\mathcal{M}_{sp}= \infty$.  The outer limit of the spall zone is clearly marked by a precipitous drop of the effective spall Mach number $\mathcal{M}_{sp}$, which crosses the $\mathcal{M}_{sp}=1$ line at $\bar{m}_k=0.135$; the actual fractured mass, established at a slightly later moment $t=0.220$, has a somewhat higher value of $\bar{m}_{sp} =0.131$. In Fig.~\ref{f:6} the respective numbers are $\bar{m}_k=0.572$ and $\bar{m}_{sp} =0.568$.

\begin{figure}[hbt]
  \centering
  \includegraphics[width=\myfigwidth]{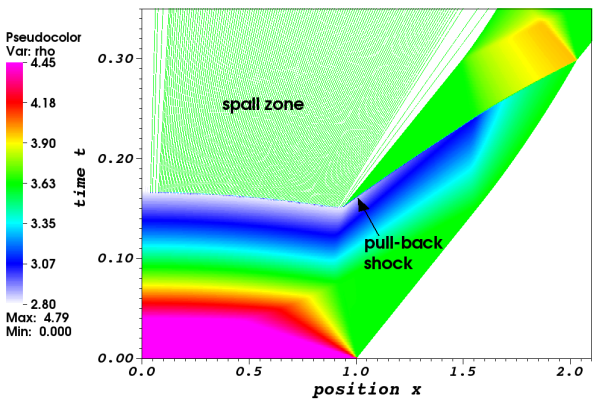}
  \caption{Same as Fig.~\ref{f:3} but for $u_p=4.0$} \label{f:5}
\end{figure}

\begin{figure}[hbt]
  \centering
  \includegraphics[width=\myfigwidth]{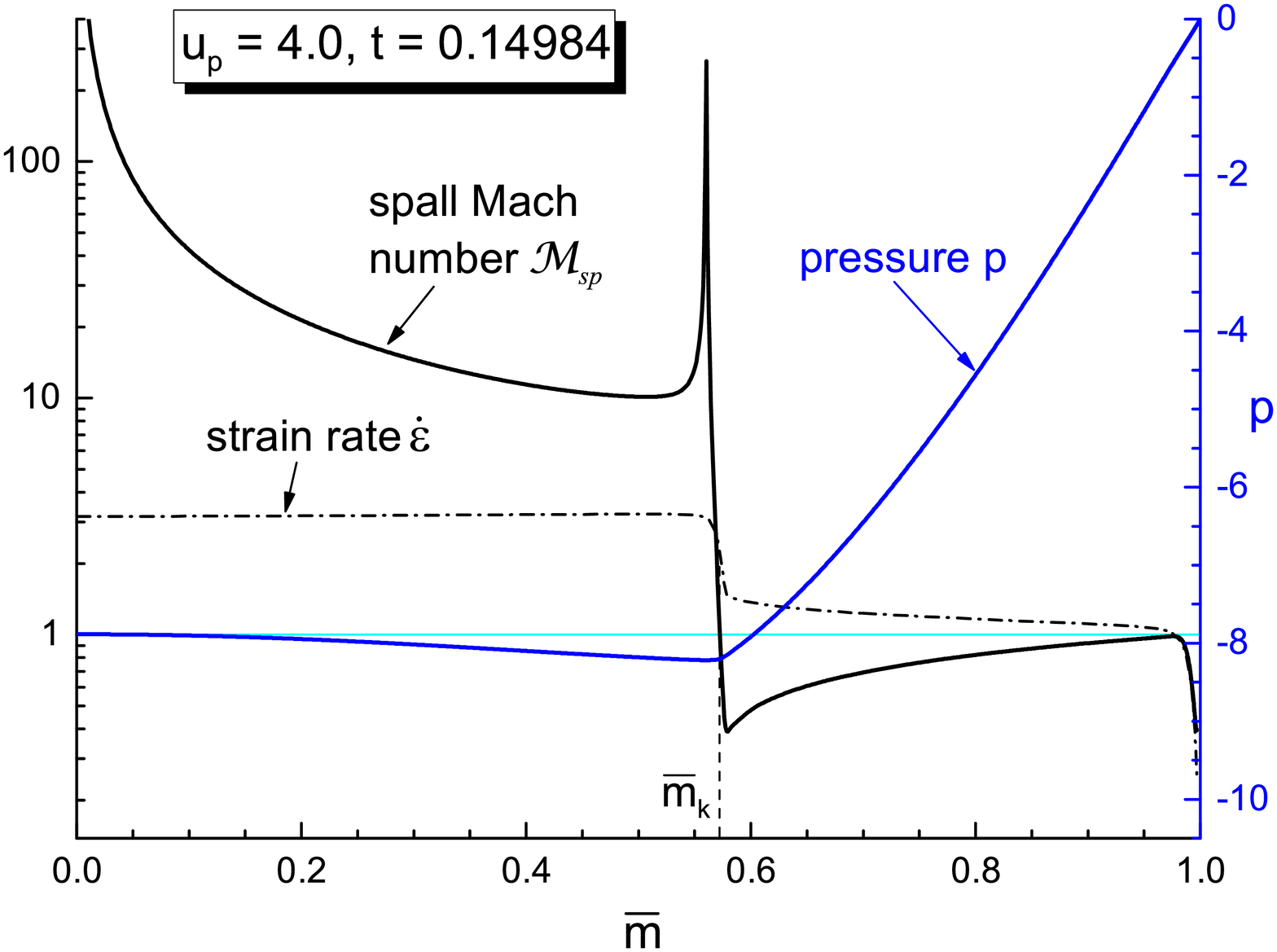}
  \caption{Same as Fig.~\ref{f:4} but for $u_p=4.0$} \label{f:6}
\end{figure}

Two more observations to be made from Figs.~\ref{f:4} and \ref{f:6} are (i)~a virtually flat pressure profile across the incipient spall zone $0< \bar{m}< \bar{m}_k$, and (ii)~a jump by about a factor 2 in the value of the strain rate $\dot{\varepsilon}$ at $\bar{m} =\bar{m}_k$. Here we must recall that the exact solution to our problem contains weak discontinuities (see problem~1 to \S~105 in~\cite[]{LL-H87}), i.e.\ discontinuities of $\partial p/\partial x$, $\partial\rho/\partial x$, $\partial u/\partial x$ that are smeared by the artificial viscosity in numerical profiles of Figs.~\ref{f:3}--\ref{f:6}. In Figs.~\ref{f:4} and \ref{f:6} the kink point $\bar{m}_k$ in the pressure profile is just such a discontinuity.  As a consequence, the exact profiles of $\mathcal{M}_{sp}$ and $\dot{\varepsilon}$ would both have jumps at $\bar{m} = \bar{m}_k$. In our simulations the threshold value of $\mathcal{M}_{sp}=1$ is bracketed with a wide margin by its pre- and post-jump values.  Due to nonlinearity of the problem, the pressure profile at $\bar{m}< \bar{m}_k$ is not exactly flat: it is monotonic with a minimum at $\bar{m}=0$ for $u_p=1.25$, and non-monotonic with a minimum at the kink point $\bar{m}_k$ for $u_p=4.0$. That is, the exact position of spall initiation may differ depending on the EOS details and the load intensity. The narrow peak of $\mathcal{M}_{sp}$ just before $\bar{m}=\bar{m}_k$ in Fig.~\ref{f:6} is caused by numerical smoothing of this corner point in the $p$ and $u$ profiles, and would not occur in the exact solution.

\subsection{On morphology of the spall zone}

The space-time image of Fig.~\ref{f:3} offers a convenient opportunity to make a comparison with the published MD simulation of a similar problem for liquid copper \cite[]{Cai_Wu.2017}. When rescaled to the critical point of copper \cite[]{Fortov_Iakubov.2006}, the parameters of our $u_p=1.25$ run are close to the MD case shown in Fig.~1 of Ref.~\onlinecite{Cai_Wu.2017}. The overall spall dynamics, the final size, and the qualitative structure of the spall zone look also quite similar. The spall strength $\sigma_{sp} \approx 11 P_{cr}$, calculated in Ref.~\onlinecite{Cai_Wu.2017}, practically coincides with the spinodal limit given in Table~\ref{t:2}. Significant differences emerge only when the structure of fractured liquid is examined in more detail.

In the results of Ref.~\onlinecite{Cai_Wu.2017}, the spall zone has a width of some 600 interatomic distances and the fractured mass splits into three liquid drops separated by voids. Generally, one would expect the MD approach to produce some statistical distribution of liquid fragments, whose typical sizes would be determined by the atomic length scale, inherent in any MD model, and by statistical properties of the microscopic initial states used in the MD code runs.

The present model, in contrast, is expected to produce a fully deterministic pattern of spall fragmentation because it is based on the equations of ideal hydrodynamics and contains no length scale to stipulate a specific fragment size. According to the arguments of Sect.~\ref{s:spf}, the entire spall zone in the exact solution to our problem should look like a single fog bubble without any remaining liquid drops because the condition $\mathcal{M}_{sp} \gg 1$ is fulfilled everywhere throughout it. The  many small droplets, observed in the spall zones of our Figs.~\ref{f:3} and \ref{f:5}, are, in fact, of spurious origin due to a short-wavelength (on the cell-size scale) numerical noise, generated by the simple numerical scheme of the DEIRA code when phase-flip pressure jumps occur at discrete time moments. As the mesh is refined, an ever more fine pattern of alternating liquid and fog cells develops, and no convergence to the exact solution is observed.

The above predicament for numerical modeling is closely related to the fact that our base model can ensure only conditional stability of the fragmentation morphology with respect to small perturbations. Because the spall Mach number (\ref{spf:Mach_sp=}) is controlled by the gradients of principal variables,
the overall fragmentation pattern (i.e.\ the number, spacing and sizes of separate liquid fragments) becomes stable with respect to small perturbations when not only the variations of $\rho$, $p$, and $\mathbf{u}$  but those of their spatial derivatives are kept sufficiently small --- hence the strong sensitivity of the fragmentation morphology to small perturbations with high wave numbers.

\begin{figure}[hbt]
  \centering
  \includegraphics[width=\myfigwidth]{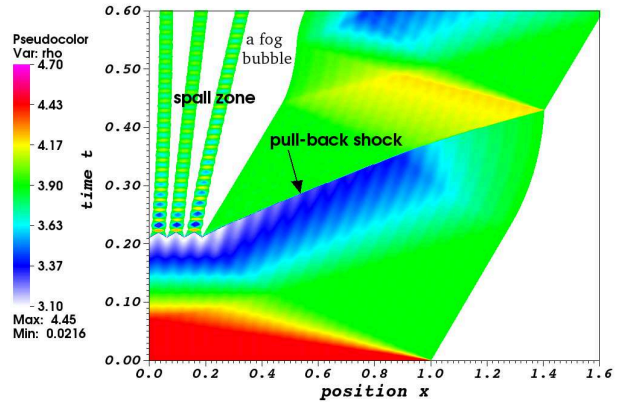}
  \caption{Same as Fig.~\ref{f:3} but calculated with the initial density perturbation (\ref{num:dp_1=})} \label{f:7}
\end{figure}

As an illustrative example, Fig.~\ref{f:7} displays the same case as in Fig.~\ref{f:3} but simulated with a perturbed initial density
\begin{equation}\label{num:dp_1=}
  \rho(0,x)= \rho_1\left[1-0.001\cos(46\pi x) \right]
\end{equation}
under the unaltered pressure $p(0,x)=p_1$. One sees that even as small as a 0.1\% density perturbation results in a dramatic change of the spall pattern: instead of a single fog bubble we have three liquid drops separated by four fog bubbles. Once well resolved by the mesh (43 cells per each drop in Fig.~\ref{f:7}), the mass and other dynamic properties of these three drops remain stable with respect to further mesh refinement.

In summary, being fully deterministic for given initial and boundary conditions, our model should predict realistic morphology of the spall zone when realistic small perturbations are included into the problem formulation.  By numerical implementation, a special care may be needed to suppress spurious short-wavelength perturbations due to finite differencing.

\section{First post-acoustic approximation for a symmetric planar spall \label{s:PA}}

\begin{figure}[hbt]
  \centering
  \includegraphics[width=\myfigwidth]{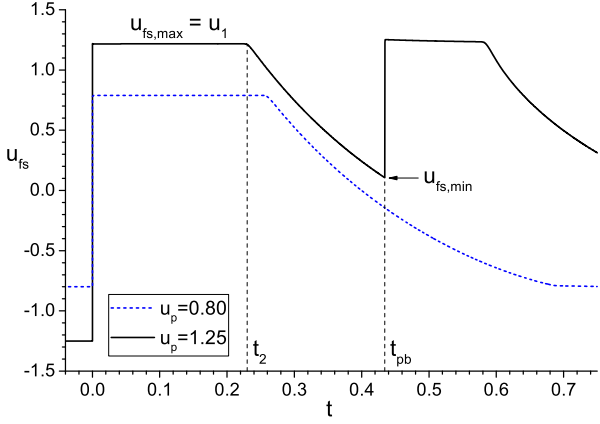}
  \caption{Time dependence of the free-surface velocity $u_{fs}(t)$ at the outer target edge $\bar{m}=1$, calculated with the DEIRA code for two cases of the initial shock load: one below ($u_p=0.8$, blue dashes) and the other above ($u_p=1.25$, black solid) the spall threshold} \label{f:8}
\end{figure}

In most plate impact experiments, the spall strength $\sigma_{th}$ is inferred from the time history of the free surface velocity $u_{fs}(t)$ measured at the outer target edge \cite[]{Antoun_Seaman.2003, Kanel_Fortov.2007}. Figure~\ref{f:8} shows the plots of $u_{fs}(t)$, calculated for two cases of the initial load in the above stated problem, one below ($u_p=0.8$) and the other above ($u_p=1.25$) the spall threshold.  Given the density $\rho_0$ and the sound speed $c_0$ of the normal state, $\sigma_{th}$ is most often evaluated in the linear acoustic approximation \cite[]{Dekel_Eliezer.1998, Antoun_Seaman.2003} from the formula
\begin{equation}\label{PA:sig_ac=}
  \sigma_{th,ac} = \frac{1}{2} \rho_0 c_0 \Delta u_{fs},
\end{equation}
where
\begin{equation}\label{PA:du_fs=}
  \Delta u_{fs} = u_{fs,max} -u_{fs,min}
\end{equation}
is the amplitude of the first dip in the $u_{fs}(t)$ profile. However,
contrary to what might be expected, formula (\ref{PA:sig_ac=}) never becomes asymptotically accurate even in the limit of an infinitely weak load because of the pb-shock attenuation: below we show its error to be 40\% for the considered case of symmetric plate collision. The situation becomes even worse when the acoustic formulae \cite[]{Dekel_Eliezer.1998,Cai_Wu.2017}
\begin{equation} \label{PA:deps_ac=}
  \dot{\varepsilon}_{sp,ac} = \frac{\Delta u_{fs}}{2c_0(t_{pb}-t_2)}, \quad
  \bar{m}_{sp,ac} = 1-\frac{c_0t_{pb}}{2h_0}
\end{equation}
for the spall strain rate $\dot{\varepsilon}_{sp}$  and the fractured mass fraction $\bar{m}_{sp}$ are invoked: the error in $\dot{\varepsilon}_{sp}$ is often a factor 3--5, while absurd negative values are easily obtained for $\bar{m}_{sp}$. In this section we go to the next order in the nonlinear terms and derive new formulae in what we call the \emph{first post-acoustic} (PA) approximation, which provide significantly better accuracy than Eqs.~(\ref{PA:sig_ac=}) and (\ref{PA:deps_ac=}) for the respective quantities.

\subsection{Characteristics chart \label{s:char}}

\begin{figure}[hbt]
  \centering
  \includegraphics[width=0.5\myfigwidth]{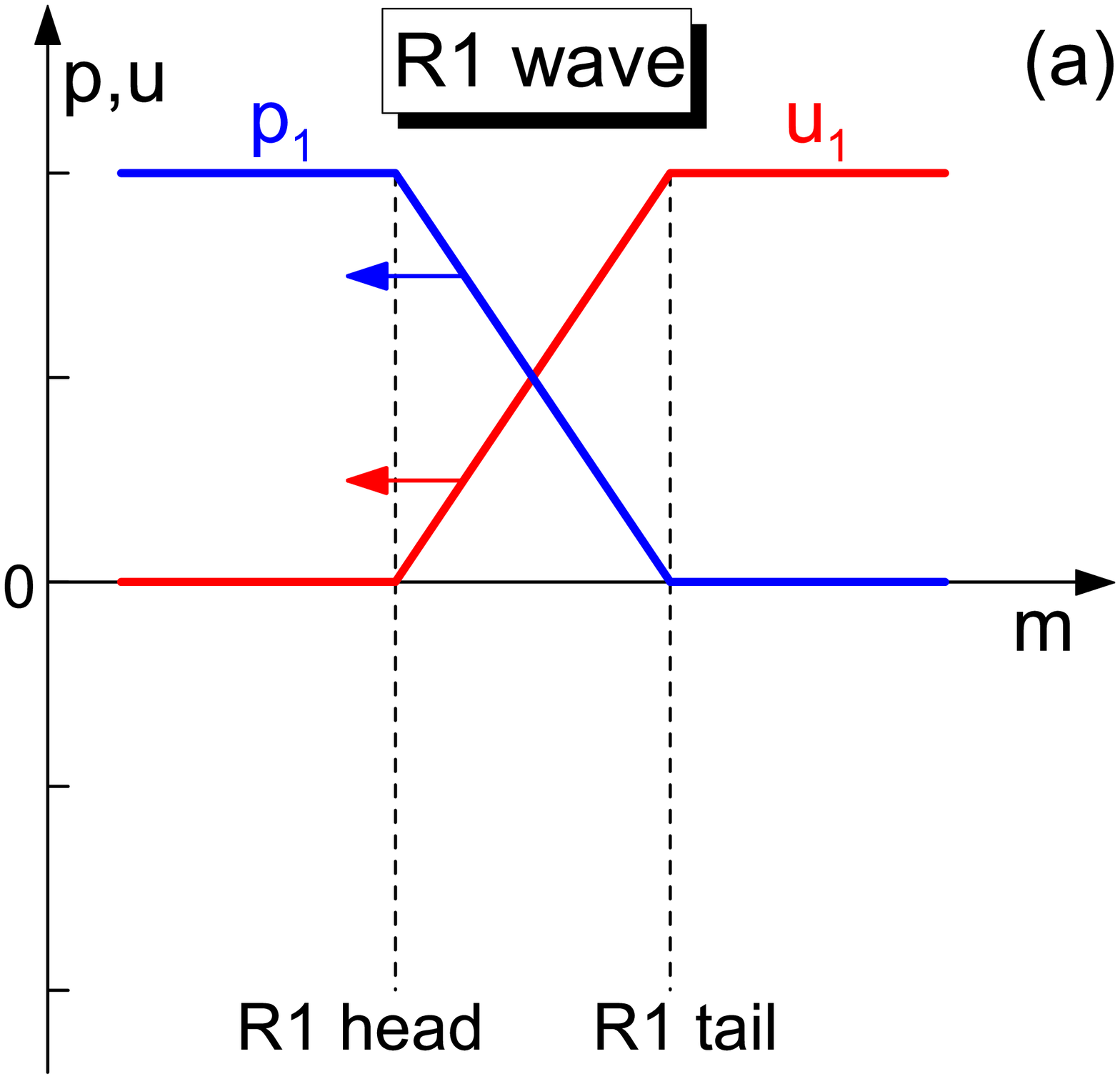}
  \includegraphics[width=0.5\myfigwidth]{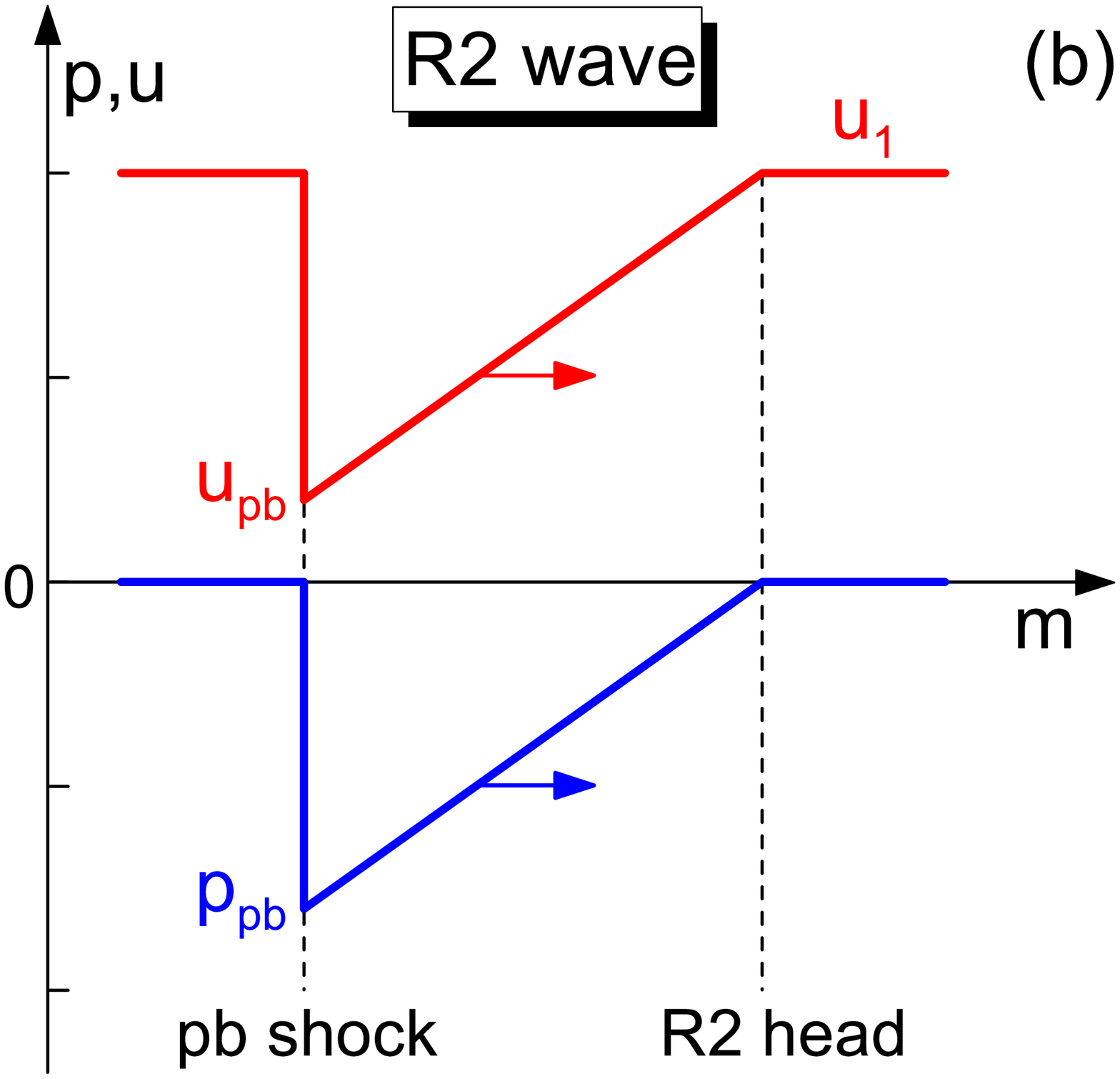}
  \caption{Schematic velocity and pressure profiles in the left-propagating R1 (a) and right-propagating R2 (b) rarefaction waves. The right-propagating R2 wave is ``propped up'' from behind by the pull-back shock} \label{f:9}
\end{figure}

As the loading shock arrives at the outer target edge $\bar{m}=1$ at $t=0$, a rarefaction wave R1 is launched in the negative direction towards the symmetry plane $\bar{m}=0$, as is schematically shown in Fig.~\ref{f:9}a. The fluid before the R1 front is at rest in the shock-compressed state $\rho_1=v_1^{-1}$, $p_1= p(\rho_1,s_1)$ (point $P$ in Fig.~\ref{f:1}); the fluid behind it moves with a constant positive velocity
\begin{equation}\label{char:u_1=}
  u_1 = \int\limits_{0}^{p_1} \frac{ dp}{\rho c} =
  \int\limits_{\rho_{01}}^{\rho_1}  c\, d\ln\rho
\end{equation}
and is in the thermodynamic state $\rho_{01}$, $p_{01} \equiv p(\rho_{01},s_1) =0$ (point $O_1$ in Fig.~\ref{f:1}). Here and below we assume that the MS EOS (\ref{p(v,tet)=})-(\ref{s(v,tet)=}) is cast in the form $p=p(\rho,s)$. So long as no new shocks emerge and the entropy remains at its post-load-shock value $s=s_1$, the pressure $p= p(\rho,s_1) = p(\rho)$ and the sound speed $c =(dp/d\rho)^{1/2} = c(\rho)$ are functions of $\rho$ only.

The key to further analysis is the behavior of characteristics for the two governing hydrodynamic equations, whose Lagrangian form in our case is
\begin{eqnarray} \label{char:dv/dt=}
  \frac{\partial v}{\partial t}- \frac{\partial u}{\partial m} &=& 0,
   \\ \label{char:du/dt=}
  \frac{\partial u}{\partial t} + \frac{\partial p}{\partial m}&=& 0,
\end{eqnarray}
where the mass coordinate $m$ and its normalized counterpart $\bar{m}$ are defined in Eq.~(\ref{num:bar_m=}). In terms of the normalized acoustic impedance
\begin{equation}\label{char:z=}
  z=z(\rho) \, \stackrel{\mbox{\scriptsize def}}{=} \,
  \frac{\rho c}{\rho_{01} c_{01}}, \quad
  z_0 = \frac{\rho_0 c_0}{\rho_{01} c_{01}},
\end{equation}
and the normalized time
\begin{equation}\label{char:bar_t=}
  \bar{t} \, \stackrel{\mbox{\scriptsize def}}{=} \,
  \frac{\rho_{01} c_{01}}{\rho_0 h_0} \,t =\frac{c_0 t}{h_0 z_0},
\end{equation}
the two families of $C^{\pm}$ characteristics on the $(\bar{m}, \bar{t})$ plane are given by
\begin{equation}\label{char:char=}
  \frac{d\bar{m}}{d\bar{t}} = \left\{ \begin{array}{ll}
  +z, \; \; C^+: \,dJ^+ \equiv du+dp/\rho c =0, \\
  -z, \; \; C^-: \,dJ^- \equiv du-dp/\rho c =0,
  \end{array}\right.
\end{equation}
where $J^{\pm}$ are the two respective Riemann invariants \cite[]{LL-H87}.

\begin{figure}[hbt]
  \centering
  \includegraphics[width=\myfigwidth]{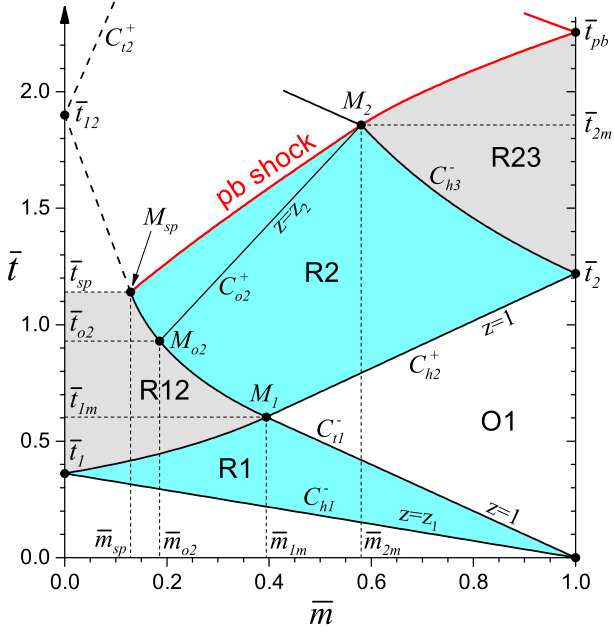}
  \caption{The pattern of characteristics in the normalized $(\bar{m},\bar{t})$ plane, depicting interaction of the release waves initiated by the shock load with $u_p=1.25$. Cyan-shaded areas show the R1 and R2 simple wave zones. Grey-shaded areas are the R12 and R23 reflection zones. Thick red curve $M_{sp}$-$M_2$-$\bar{t}_{pb}$ is the trajectory of the pull-back shock } \label{f:10}
\end{figure}

Figure~\ref{f:10} shows the pattern of $C^{\pm}$ characteristics, representing the solution to our problem in the normalized $(\bar{m},\bar{t})$ plane. The R1 wave starts at point~1 with $(\bar{m},\bar{t}) =(1,0)$ as a centered $C^-$ fan between the leading $C_{h1}^-$ and the trailing $C_{t1}^-$ lines. As each of these $C^-$ characteristics reaches the symmetry plane $\bar{m}=0$, it is reflected and continues as a $C^+$ characteristic, i.e.\ $C_{h2}^+$ is the reflected continuation of $C_{h1}^-$. Where the R1 fan borders a region of constant flow, it is a \emph{simple} wave with all the $C^-$ characteristics being straight. In Fig.~\ref{f:10} the R1 simple wave (shaded cyan) is confined to a partially curvilinear triangle bounded by the contour 1-$\bar{t}_1$-$M_1$-1, inside which $J^+$ is constant. In the left reflection zone R12 (shaded grey), bounded by the contour $\bar{t}_1$-$M_1$-$\bar{t}_{12}$-$\bar{t}_1$, the flow is not a simple wave, and all the $C^{\pm}$ characteristics are curved.

As the R1 wave is reflected from the left edge $\bar{m}=0$, it becomes the R2 simple wave (shown schematically in Fig.~\ref{f:9}b) beyond the R12 reflection zone. If the pb shock is launched from point $M_{sp}$, lying on the $C^-_{t1}$ characteristic, the R2 simple-wave zone (shaded cyan in Fig.~\ref{f:10}) is bounded by the contour $M_1$-$M_{sp}$-$M_2$-$\bar{t}_2$-$M_1$, where $M_{sp}$-$M_2$ is the curved trajectory of the pb shock and $\bar{t}_2$-$M_2$ is a curved segment of the $C_{h3}^-$ characteristic, being the continuation of $C_{h2}^+$ upon its reflection off the free surface $\bar{m}=1$. The curvilinear triangle $\bar{t}_2$-$M_2$-$\bar{t}_{pb}$ (shaded grey), where $\bar{t}_{pb}$ is the time of the pb-shock arrival at $\bar{m}=1$, encompasses the right reflection zone R23. In the R2 zone, all the $C^+$ characteristics are straight --- like the $C^+_{2o}$ characteristic passing through $M_2$. In the absence of spall, the R1 trailing characteristic $C^-_{t1}$ is reflected from the $\bar{m}=0$ edge at $\bar{t}_{12}$ and continues as the $C^+_{t2}$ straight line, which replaces the pb shock as the downstream boundary of the R2 simple wave.

The O1 zone inside the rectilinear isosceles triangle 1-$M_1$-$\bar{t}_2$ is the region of constant flow with $u=u_1$, $\rho= \rho_{01}$, $c=c_{01}$, $p=0$. The times $t_2$ and $t_{pb}$ in Fig.~\ref{f:8}, which mark the beginning and the end of the depression in the $u_{fs}(t)$ curve, are the non-normalized counterparts of $\bar{t}_2$ and $\bar{t}_{pb}$ in Fig.~\ref{f:10}. Note that all the head and tail characteristics, plotted in Fig.~\ref{f:10}, are also the trajectories of weak flow discontinuities.

\subsection{Release isentrope in the PA approximation \label{s:isPA}}

To construct the desired PA approximation, we need certain information about the isentrope $p(\rho)$ (curve $PO_1S$ in Fig.~\ref{f:1}), passing through the zero-pressure state $O_1$ at $\rho=\rho_{01}$. We begin by assuming that, in addition to $\rho_{01}$ and $c_{01}=c(\rho_{01})$, one knows from shock experiments the coefficient $b$ in the widely used linear approximation
\begin{equation}\label{isPA:D=}
  D =c_{01} +bu'
\end{equation}
to the compression Hugoniot with the initial state at $O_1$; here $D$ and $u'$ are, respectively, the shock-front and the post-shock particle velocities relative to the fluid in zone O1, which itself moves with velocity $u_1$ in our coordinate system. Having applied the mass and momentum balance relations across the shock front, we can convert (\ref{isPA:D=}) into the parametric representation of the Hugoniot
\begin{eqnarray} \label{isPA:p_H=}
  p_H(u') &=& \rho_{01} u'(c_{01}+bu'), \\ \label{isPA:rho_H=}
  \rho_H(u') &=& \rho_{01} \frac{c_{01}+bu'}{c_{01}+(b-1)u'}.
\end{eqnarray}

\begin{figure}[hbt]
  \centering
  \includegraphics[width=\myfigwidth]{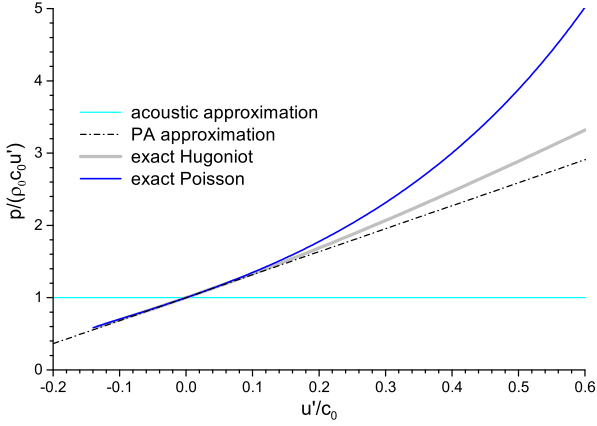}
  \caption{PA approximation to the release isentrope (dash-dotted), passing through the normal state $\rho_0$, $p_0=0$, versus particle velocity increment $u'$. Shown also are the exact Hugoniot (solid gray) and Poisson (solid blue) adiabats calculated from Eqs.~(\ref{p(v,tet)=})--(\ref{s(v,tet)=})} \label{f:11}
\end{figure}

If we consider now an arbitrary isentropic simple wave, bordering zone O1 in Fig.~\ref{f:10} and propagating in the positive direction, we can similarly parametrize the isentrope $p(\rho)$ in terms of the particle velocity increment $u'$ by integrating the equation
\begin{equation}\label{isPA:dp/du'=}
  du' = dp/(\rho c) = c\, d\ln\rho,
\end{equation}
which expresses the fact that $dJ^-=0$ throughout both the O1 zone and the bordering simple-wave regions. In doing this, we can simply put $p(u')=p_H(u')$ because the Hugoniot and the Poisson adiabats have a second-order contact at the common origin point, which implies $dp/du'= dp_H/du'$ and $d^2p/du'^2 = d^2p_H/du'^2$ at $u'=0$.  Having applied this argument to our R1 and R2 simple rarefaction waves, we obtain the following PA approximation to the required isentrope
\begin{eqnarray} \label{isPA:p=}
  p(\bar{u}') &=& \rho_{01} c_{01}^2\, \left(\bar{u}' +b\bar{u}'^2 \right),
  \\ \label{isPA:z=}
  z(\bar{u}') &=& 1+ 2b\bar{u}',
\end{eqnarray}
where
\begin{equation}\label{isPA:bar_u'=}
  \bar{u}' = \frac{u'}{c_{01}}, \quad u'= \left\{ \begin{array}{ll}
  u_1-u >0, & \mbox{R1 wave}, \\
  u-u_1 <0, & \mbox{R2 wave}.
  \end{array} \right.
\end{equation}
Below, we refer to the second terms on the right-hand sides of Eqs.~(\ref{isPA:p=}) and (\ref{isPA:z=}) as the \emph{pu-nonlinearity corrections}. They characterize the nonlinearity of the pertinent EOS and are proportional to the value of the fundamental gasdynamic derivative \cite[]{Thomson1971}
\begin{equation}\label{isPA:Gam=}
  \Gamma = \frac{c^4}{2v^3} \left(\frac{\partial^2 v}{\partial p^2}\right)_s =
  \frac{v^3}{2c^2} \left(\frac{\partial^2 p}{\partial v^2}\right)_s =2b
\end{equation}
in the state where $\bar{u}'=0$. The accuracy of the PA approximation to the GWEOS isentrope, passing through the normal state with parameters from Table~\ref{t:1}, is illustrated in Fig.~\ref{f:11}, where the essentially positive dimensionless ratio $p/(\rho_0 c_0 u')$ is plotted versus $\bar{u}'$. The error in $p$ stays below 5\% for $\left|\bar{u}'\right| \lesssim 0.15$, but exceeds 30\% for $\bar{u}' \gtrsim 0.5$.

If Eqs.~(\ref{isPA:p=}), (\ref{isPA:z=}) are deemed to be exact along a considered isentrope, they yield
\begin{eqnarray}
  \rho(\bar{u}') &=& \rho_{01}\left[1-(2b)^{-1} \ln(1+2b\bar{u}')\right]^{-1}
  \nonumber \\ \label{isPA:rho=}
  &=& \rho_{01}\left[1+\bar{u}'-(b-1)\bar{u}'^2 + \ldots\right], \\
  c(\bar{u}') &=& c_{01}(1+2b\bar{u}') \left[1-(2b)^{-1} \ln(1+2b\bar{u}')\right]^{-1}
  \nonumber \\ \label{isPA:c=}
  &=& c_{01} \left[1+(2b-1)\bar{u}' -b\bar{u}'^2 +\ldots \right].
\end{eqnarray}
Because neither $\rho$ nor $z$ can be negative, the applicability of the above formulae is in any case limited to the range
\begin{equation}\label{isPA:<=u'<=}
  -1/2b < \bar{u}' <  \left( e^{2b}-1\right)/2b,
\end{equation}
which is usually sufficient for application to spall experiments. In fact, if separate values of $\rho(u')$ or $c(u')$ are needed, one can justifiably restrict oneself to the zero- and first-order terms in powers of $\bar{u}'$.

\subsection{Evaluation of the strain rate \label{s:sr}}

\begin{figure}[hbt]
  \centering
  \includegraphics[width=\myfigwidth]{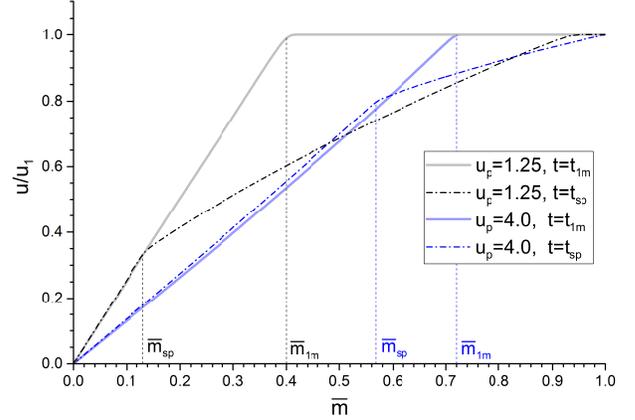}
  \caption{Normalized velocity profiles at the moment $t_{1m}$ of R1 reflection  and at the moment $t_{sp}$ of spallation, calculated with the DEIRA code for the two reference cases of $u_p=1.25$ (black) and $u_p=4.0$ (blue)} \label{f:12}
\end{figure}

Introduction of nonlinear terms opens up a possibility to obtain an a priori estimate of the strain rate at spallation without invoking the $u_{fs}(t)$ curve --- the possibility not available in the acoustic approximation. Our estimate is based on the assumption of time-independent linear velocity profile
\begin{equation}\label{sr:LU-app=}
  \frac{\partial u(t,m)}{\partial m} \approx
  \frac{\partial u(t_{1m},m)}{\partial m} \approx   \frac{u_1}{m_{1m}}
\end{equation}
throughout the reflection zone R12  --- the LU approximation; here the \emph{mid-reflection} time $t_{1m}$ is defined as the moment when the characteristics $C^-_{t1}$ and $C^+_{h2}$ in Fig.~\ref{f:10} intersect at point $M_1$, and the zone O1 of constant flow behind R1 has its maximum extension $m_0-m_{1m}$ along the $m$ coordinate. Figure~\ref{f:12} confirms that this approximation remains quite accurate even in the highly nonlinear case of $u_p=4.0$.

The LU approximation (\ref{sr:LU-app=}) can be supported by the observation that the piece-wise linear profile of $u$, displayed in Fig.~\ref{f:9}a, is in fact the exact solution of the nonlinear Euler equation for $u(t,m)$, obtained by substituting Eq.~(\ref{isPA:p=}) into Eq.~(\ref{char:du/dt=}); in this solution the width $\Delta m =(\rho_1c_1-\rho_{01}c_{01})t$ of the R1 wave grows linearly in time. When the reflection of such a linear in $m$ rarefaction pulse from the $m=0$ boundary is treated in the acoustic limit of $c_1-c_{01} \ll c_{01}$, both approximate equalities in (\ref{sr:LU-app=}) become exact. Also, the fluid state at mid-reflection is then everywhere uniform with $p=0$ and $\rho= \rho_{01}$. Note that the slope (\ref{sr:LU-app=}) remains finite as both $u_1$ and $m_{1m}$ become infinitely small in the limit of $u_p \to 0$, $c_1\to c_{01}$.

To evaluate the mid-reflection coordinate $\bar{m}_{1m}= m_{1m}/m_0$, we use the obvious fact that the integral $\oint d\bar{t}=0$ along any closed contour in the $(\bar{m},\bar{t})$ plane. Then we combine this circulation rule with Eq.~(\ref{char:char=}) and apply it to the perimeter of the R1 simple-wave zone. Whereas integration along straight characteristic segments, on which all flow parameters remain constant, is trivial, for any curved segment --- like $\bar{t}_1$-$M_1$ in Fig.~\ref{f:10} --- we assume the acoustic impedance to be approximately constant and equal to its arithmetic mean $z= (z_a+z_b)/2$ between the end values $z_a$ and $z_b$. Thus, the condition of zero $\oint d\bar{t} = \oint (d\bar{m}/d\bar{t})^{-1} d\bar{m}$ along the triangular contour 1-$\bar{t}_1$-$M_1$-1 yields the equation
\begin{equation}\label{sr:Eq1=}
  z_1^{-1} +2(1+z_1)^{-1} \bar{m}_{1m} - (1-\bar{m}_{1m})=0,
\end{equation}
from which we get
\begin{equation}\label{sr:bm_1m_PA=}
  \bar{m}_{1m} = \frac{z_1-z_1^{-1}}{z_1+3} =\frac{2\beta_1(1+\beta_1)}{(2+\beta_1)(1+2\beta_1)},
\end{equation}
where, according to (\ref{isPA:z=}),
\begin{equation}\label{sr:z_1=}
  z_1=1+2\beta_1, \quad \beta_1=bu_1/c_{01}.
\end{equation}
Note that the acoustic limit corresponds to $\beta_1 \ll 1$.

Now, with $\partial u/ \partial m$ known from Eq.~(\ref{sr:LU-app=}), the strain rate (\ref{spf:deps=}) can be calculated from Eq.~(\ref{char:dv/dt=}) as
\begin{equation}\label{sr:deps=}
  \dot{\varepsilon} =\rho\frac{\partial v}{\partial t} = \rho\frac{\partial u}{\partial m}.
\end{equation}
Before finalizing our estimate of $\dot{\varepsilon}$, we recall that the spatial derivatives of $u$ and $p$ are discontinuous on the $C^-_{t1}$ characteristic --- as is manifested by the kinks at $\bar{m}= \bar{m}_{sp}$  on the velocity profiles in Fig.~\ref{f:12}. Hence, we have to choose between the two different values $(\partial u/ \partial m)_\pm$ obtained on the two sides $m= m_{sp} \pm 0$ of the spall boundary. Because it is the material at $m<m_{sp}$ which undergoes spall fracture, we make the logical choice of $(\partial u/ \partial m)_- =u_1/m_{1m}$. In contrast, the often used acoustic formula (\ref{PA:deps_ac=}) represents the value $(\partial u/ \partial m)_+$ in the unfractured material, which is exactly one half of $(\partial u/ \partial m)_-$ in the acoustic limit. Thus, having combined Eqs.~(\ref{sr:LU-app=})--(\ref{sr:deps=}), we obtain the following PA estimates for the strain rates at mid-reflection,
\begin{equation}\label{sr:deps_1m_PA=}
  \dot{\varepsilon}_{1m} = \frac{\rho_{01}u_1}{m_{1m}}=
  \frac{c_0}{h_0z_0} \frac{(2+\beta_1)(1+2\beta_1)}{2b(1+\beta_1)},
\end{equation}
and at spallation
\begin{equation}\label{sr:deps_sp=}
  \dot{\varepsilon}_{sp} = \frac{\rho_{sp}}{\rho_{01}}\, \dot{\varepsilon}_{1m} = \left(1-\left|\bar{u}'_{sp} \right| \right)
  \dot{\varepsilon}_{1m}.
\end{equation}
The small correction $\left|\bar{u}'_{sp} \right| \simeq \sigma_{th}/\rho_0c_0^2$, accounting for the fluid expansion between the $t_{1m}$ and $t_{sp}$ moments, is typically a few percent even for the highest theoretical values of $\sigma_{th}$ \cite[]{Mayer_Mayer2015, Cai_Wu.2017, Mayer_Mayer2020} and can usually be neglected; the PA formula for $|\bar{u}'_{sp}|$  is given in the next section.

If time $t_2$ is known from the measurements of the free surface velocity $u_{fs}(t)$, one readily obtains an alternative a posteriori LU estimate
\begin{eqnarray}\label{sr:bm_m1_LU=}
  \bar{m}_{1m} &=& 1-\frac{1}{2} \frac{c_0t_2}{h_0z_0},
  \\ \label{sr:deps_1m_LU=}
  \dot{\varepsilon}_{1m} &=& \frac{c_0}{h_0}
  \frac{u_1/c_{01}}{z_0-c_0t_2/2h_0},
\end{eqnarray}
which is a direct consequence of the LU approximation (\ref{sr:LU-app=}) and of the obvious fact that $t_{1m}=t_2/2$. Note that, having the dimension of $t^{-1}$, the strain rate $\dot{\varepsilon}_{1m}$ in Eqs.~(\ref{sr:deps_1m_PA=}) and (\ref{sr:deps_1m_LU=}) scales in inverse proportion to the initial target thickness $h_0$.

Regarding practical application of Eqs.~(\ref{sr:deps_1m_PA=}) and (\ref{sr:deps_sp=}), we will distinguish two subcases of the PA approximation depending on the available thermodynamic information for the material in question, namely, a cruder PA0 version and a more accurate PA1 variant:
\begin{equation}
\nonumber 
  \begin{array}{lll}
  \mbox{PA0:} & \rho_0, c_0, b_0  \rightarrow  & \mbox{known,} \\
              &\rho_{01},c_{01},b,u_1 \rightarrow &
              \mbox{replaced by } \rho_0,c_0,b_0,u_p; \\[\jot]
  \mbox{PA1:} & \rho_0, c_0, \rho_{01},c_{01},b,u_1 \rightarrow &
                                             \mbox{all known.}
  \end{array}
\end{equation}
In the PA0 version, only the normal-state values of $\rho_0$, $c_0$, $b_0$ are assumed to be known, while for the PA1 case one needs also the parameters of the $p=0$ state $O_1$ on the release isentrope behind the initial load shock. Analogously, the LU approximation is subdivided into a cruder LU0 and a more accurate LU1 versions.

\begin{table}[hbt]
\caption{Comparison of the strain rates, evaluated in different versions of the PA and LU approximations, with those obtained in numerical simulations with the DEIRA code}
\label{t:3}
\begin{ruledtabular}
\begin{tabular}{lcccccc}
       & DEIRA & PA1 & PA0 & LU1 & LU0 & ac \\
  \hline
   \multicolumn{7}{c}{$u_p=1.25$, $\Delta u_{fs}=1.111$} \\
  \hline
  $\bar{m}_{1m}$ & 0.40 & 0.350 & 0.356 & 0.401  & 0.392 & -- \\
  $(h_0/c_0)\dot{\varepsilon}_{1m}$  & 0.51 & 0.578 & 0.587 & 0.505  & 0.532 & -- \\
  $(h_0/c_0)\dot{\varepsilon}_{sp}$  & 0.41 & 0.500 & 0.508 & 0.437 & 0.461  & 0.086 \\
 [8pt] \hline
   \multicolumn{7}{c}{$u_p=4.0$, $\Delta u_{fs}=1.063$} \\
  \hline
  $\bar{m}_{1m}$ & 0.72 & 0.603 & 0.612 & 0.717  & 0.633 & -- \\
  $(h_0/c_0)\dot{\varepsilon}_{1m}$  & 0.89 & 0.963 & 1.090 & 0.811  & 1.054 & -- \\
  $(h_0/c_0)\dot{\varepsilon}_{sp}$  & 0.66 & 0.812 & 0.947 & 0.683 & 0.915  & 0.125 \\
\end{tabular}
\end{ruledtabular}
\end{table}

Table~\ref{t:3} compares the values of $\bar{m}_{1m}$, $\dot{\varepsilon}_{1m}$, and $\dot{\varepsilon}_{sp}$, evaluated in different versions of the PA and LU approximations, with those obtained in the numerical DEIRA runs for the two reference cases of the shock load; the $ac$ column lists the values given by the acoustic formula~(\ref{PA:deps_ac=}); the strain rate is given in terms of the scale-invariant dimensionless product $(c_0/h_0)\dot{\varepsilon}$. All the required EOS parameters are taken from Tables~\ref{t:1} and \ref{t:2}; the values of $\Delta u_{fs}$, used to calculate $\dot{\varepsilon}_{sp,ac}$ and $\bar{u}'_{sp}$, as well as the times $t_2$ are inferred from the DEIRA produced plots of $u_{fs}(t)$ as shown in Fig.~\ref{f:8} for the case of $u_p=1.25$. The numbers in Table~\ref{t:3} demonstrate that the PA formulae (\ref{sr:deps_1m_PA=}), (\ref{sr:deps_1m_PA=}) provide a robust estimate of the strain rate at spallation with an accuracy of about 15--30\%, which is not sensitive to the loading shock strength. The conventional acoustic formula (\ref{PA:deps_ac=}), on the contrary, tends to underestimate $\dot{\varepsilon}_{sp}$ by a significant factor of about 3--5, as was already noticed in MD simulations \cite[]{Luo_An.2009, Cai_Wu.2017}. If one can reliably measure the arrival time $t_2$ of the R2 rarefaction wave, Eqs.~(\ref{sr:deps_sp=})--(\ref{sr:deps_1m_LU=}) yield even more accurate values of $\bar{m}_{1m}$ and $\dot{\varepsilon}_{sp}$, which confirms the high accuracy of the LU approximation (\ref{sr:LU-app=}).

\subsection{Evaluation of the spall strength \label{s:ss}}

Similar to the acoustic formula (\ref{PA:sig_ac=}), our PA estimate of the spall strength is based on the measured free-surface velocity pullback $\Delta u_{fs}$. We begin by arguing that the outer boundary $\bar{m}_{sp}$ of the spall zone, represented by the pb-shock starting point $M_{sp}$  in Fig.~\ref{f:10}, should generally lie on the R1 trailing characteristic $C^-_{t1}$, where $\partial p/\partial m$ and  $\partial u/\partial m$ have a jump. Firstly, we note that everywhere above and to the right of the $M_1$-$\bar{t}_{12}$ segment of the $C^-_{t1}$ curve, where the pressure is negative in the R2 simple-wave zone, we have
\begin{equation}\label{ss:Mach_sp=}
  \mathcal{M}_{sp} = c/D_{pb} < 1.
\end{equation}
This inequality, obtained by substituting Eq.~(\ref{sr:deps=}) and the relation $dp=\pm\rho c\, du$ into the definition (\ref{spf:Mach_sp=}) of the effective spall Mach number $\mathcal{M}_{sp}$, implies that the pb shock must originate either from the $C^-_{t1}$ characteristic or from the interior of the R12 reflection zone. On the other hand, we know that in the acoustic limit $\beta_1 \ll 1$ it cannot start inside the R12 zone because in this limit  $\partial p/\partial m \to 0$ and $\mathcal{M}_{sp} \to \infty$ everywhere inside R12. This proves that at least for not too large values of $u_p$ and $\beta_1$ the pb shock can only start from the $C^-_{t1}$ characteristic. In strongly nonlinear cases with $\beta_1 \gtrsim 1$, this fact is confirmed by numerical simulations (at least for the present EOS), as is shown in Fig.~\ref{f:6}.

With $M_{sp}$ lying on the $C^-_{t1}$ characteristic above the $M_1$ point, we can use the analytic properties of the R2 simple wave, represented by Eqs.~(\ref{isPA:p=})--(\ref{isPA:c=}), to relate $\sigma_{sp}$ to $\Delta u_{fs}$. The simple acoustic formula (\ref{PA:sig_ac=}) is based on the premise that the negative-pressure spall pulse (see Fig.~\ref{f:9}b), created at $M_{sp}$ with the initial velocity decrement
\begin{equation}\label{ss:u'_sp=}
  u'_{sp} \equiv c_{01}\bar{u}'_{sp} \, \stackrel{\mbox{\scriptsize def}}{=} \, u(t_{sp},m_{sp})-u_1 <0,
\end{equation}
propagates without attenuation down to the free surface, where the well-known \cite[\S~11, Ch.~XI]{Zeldovich_Raizer2012} velocity doubling rule $\Delta u_{fs}= -2u'_{sp}$ applies; then (\ref{PA:sig_ac=}) follows directly from Eq.~(\ref{isPA:p=}) in the limit of $b|\bar{u}'| \ll 1$.

In the PA approximation, we take into account the attenuation of the spall pulse, caused by the decreasing amplitude of the pb shock as it propagates up the rising density profile of the R2 simple wave. For this, we trace the curvilinear pb-shock trajectory up to its intersection with the right-reflected characteristic $C^-_{h3}$ at point $M_2$ and apply the doubling rule
\begin{equation}\label{ss:Du_fs=}
  \Delta u_{fs}= -2u'_2
\end{equation}
to the velocity decrement
\begin{equation}\label{ss:u'_2=}
  u'_2 \equiv c_{01}\bar{u}'_2 \, \stackrel{\mbox{\scriptsize def}}{=} \, u(t_{2m},m_{2m})-u_1 <0.
\end{equation}
at that point, which has coordinates $(\bar{m}_{2m},\bar{t}_{2m})$ in Fig.~\ref{f:10}. Having introduced the \emph{pb-attenuation correction}
\begin{equation}\label{ss:del_pb-def=}
  \delta_{pb} \, \stackrel{\mbox{\scriptsize def}}{=} \,
   u'_{sp}/u'_2-1
\end{equation}
and invoking Eq.~(\ref{isPA:p=}), we obtain the following PA formula for the spall strength
\begin{eqnarray}
  \sigma_{sp} &=& -p(\bar{u}'_{sp}) = \rho_{01} c_{01}^2 |\bar{u}'_{sp}|
  \left(1-b|\bar{u}'_{sp}| \right)
  \nonumber \\ \label{ss:sig_sp=}
  &=& \frac{1}{2} \rho_0 c_0 \Delta u_{fs}\, z_0^{-1}
  (1+\delta_{pb}) \left[1-\beta_2(1+\delta_{pb}) \right],
\end{eqnarray}
where
\begin{equation}\label{ss:bet_2=}
  \beta_2 =  b|\bar{u}'_2| =
  b\frac{\Delta u_{fs}}{2 c_{01}},\quad
  |\bar{u}'_{sp}|= (1+\delta_{pb})\frac{\Delta u_{fs}}{2 c_{01}}.
\end{equation}
The attenuation correction
\begin{equation}\label{ss:del_pb=}
  \delta_{pb} = \frac{2(1-2\beta_2)}{\omega +\left[\omega^2-
  2\beta_2(1-2\beta_2)(3\mu+1) \right]^{1/2}}
\end{equation}
is found as the solution of the quadratic equation (\ref{A:Eq-del_pb=}), where
\begin{equation}\label{ss:mu,omega=}
  \mu=\frac{1+\beta_1}{1+\frac{3}{2}\beta_1} \, \frac{1-\frac{3}{2}\beta_2}{1-\frac{1}{2}\beta_2}, \quad \omega=\frac{1}{2}+\mu
  \left(2-3\beta_2\right).
\end{equation}
Note that formula (\ref{ss:sig_sp=}) may be considered as a generalization of the simple extrapolation of the Hugoniot into the negative-pressure region, advocated in Ref.~\onlinecite{Kanel2010}, which takes a more consistent account of the first-order nonlinear terms. For the fractional mass of the spall zone we obtain
\begin{equation}\label{ss:m_sp=}
  \bar{m}_{sp} =\bar{m}_{1m}\left(1- \frac{|u'_{sp}|}{u_1}\right) =
  \bar{m}_{1m}\left[1- \frac{(1+\delta_{pb})\Delta u_{fs}}{2u_1}\right],
\end{equation}
where $\bar{m}_{1m}$ is given by Eq.~(\ref{sr:bm_1m_PA=}). The derivation of Eqs.~(\ref{ss:del_pb=})--(\ref{ss:m_sp=}) is detailed in Appendix~\ref{s:A}.

\begin{figure}[hbt]
  \centering
  \includegraphics[width=0.75\myfigwidth]{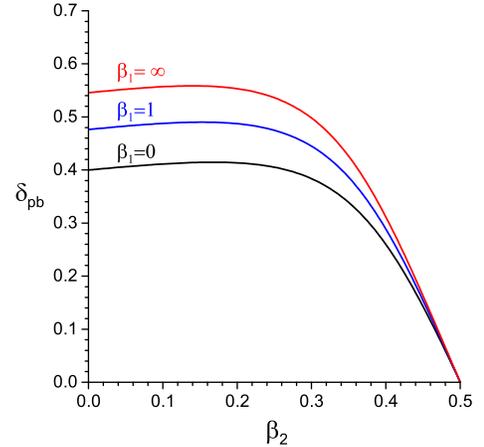}
  \caption{The pb-attenuation correction $\delta_{pb}= \delta_{pb}(\beta_1,\beta_2)$ as a function of $\beta_2$ for three selected values of $\beta_1$} \label{f:13}
\end{figure}

Although the effect of the pb-shock attenuation is fairly well known and has been discussed in literature \cite[]{Antoun_Seaman.2003,Cai_Wu.2017}, the explicit formula (\ref{ss:del_pb=}) for the corresponding correction $\delta_{pb}$ is an important new result of this paper. For the particular considered problem of symmetric plate collision, this correction is a function of two dimensionless parameters $\beta_1$ and $\beta_2$ representing (in relative terms), respectively, the intensity of the initial load and the amplitude of the spall pulse. Figure~\ref{f:13}, which displays the behavior of $\delta_{pb}(\beta_1,\beta_2)$ in the allowed range of $0 \leq \beta_2 \leq 0.5$, reveals that for $\beta_2 \lesssim 0.3$ the pb-attenuation correction stays within a narrow range $0.4\leq \delta_{pb}< 0.56$ for any value of $\beta_1>0$, i.e.\ for arbitrary velocity of the impactor plate. This observation is important for applications because in real experiments the limit of $\beta_2 \approx 0.3$ is hardly ever exceeded.

Of special interest is the seemingly paradoxical result that in the limit of $\beta_1,\beta_2 \to 0$ the pb-attenuation correction does not vanish but approaches the value
\begin{equation}\label{ss:del_00=}
  \delta_{00} \equiv \delta_{pb}(0,0) = \frac{2}{5}
\end{equation}
that is not much lower than the upper limit of $\delta_{pb,max} \approx 0.56$. The latter is explained by the fact that for $\beta_1 \sim \beta_2 \ll 1$ the small amplitude of the pb shock has the same order of magnitude as the difference between its propagation speed and the local sound velocity. In this sense, attenuation of the pb shock is qualitatively similar to the attenuation of weak saw-tooth-like pulses \cite[\S~102]{LL-H87} calculated in the first-order nonlinear approximation and controlled by the value of the fundamental gasdynamic derivative $\Gamma$ defined in Eq.~(\ref{isPA:Gam=}). Remarkably, in our case $\delta_{00}$ turns out to be a universal constant which depends neither on $\Gamma$ nor on target parameters. Its value can be calculated independently and more rigorously by direct integration of the equation (\ref{A:dm_pb/dt=}) for the pb-shock trajectory in the asymptotic limit of $\beta_1,\beta_2 \to 0$, which yields $\delta_{00} =\sqrt{2}-1$ that is practically indistinguishable from $\delta_{00} =0.4$ obtained from Eq.~(\ref{ss:del_pb=}).

In our PA formula (\ref{ss:sig_sp=}) for the spall strength, the nonlinear corrections appear as three different terms, namely, the modified (compared to the normal state) acoustic impedance $\rho_{01}c_{01} =z_0^{-1} \rho_0c_0$, the pb-attenuation correction $\delta_{pb}$, and the negative pu-nonlinearity correction  $-\beta_2(1+\delta_{pb})$. The relative importance of these correction terms can be judged from the data in Tables~\ref{t:2} and \ref{t:4}. The negative impedance correction $z_0^{-1}-1$ is small (a few percent) for moderate loads near the spall threshold, but becomes significant for highly nonlinear cases with large $u_p$. The positive pb-attenuation correction $\delta_{pb}$ is always significant and must never be ignored, at least not in the experiments where the flyer and the target plates have comparable masses.

\begin{table}[hbt]
\caption{ Comparison of the PA and acoustic (ac) approximations with the exact (EOS or DEIRA) results for the spall strength $\sigma_{th}$ and the spalled mass fraction $\bar{m}_{sp}$, calculated for the two reference values of $u_p$. The values of $\Delta u_{fs}$ are taken from the $u_{fs}(t)$ curves produced by the DEIRA runs}
\label{t:4}
\begin{ruledtabular}
\begin{tabular}{lllll}
       & EOS/DEIRA & PA1 & PA0 &  ac \\
  \hline
   \multicolumn{5}{c}{$u_p=1.25$, $\Delta u_{fs}=1.111$} \\
  \hline
  $\beta_1$     & -- & 0.6442 & 0.6614 & -- \\
  $\beta_2$     & -- & 0.2941 & 0.2939 & -- \\
  $\delta_{pb}$ & -- & 0.4370 & 0.4378 & -- \\
  $\sigma_{sp}$ & 11.15 (EOS) & 10.65 & 10.80   & 13.01 \\
  $\bar{m}_{sp}$ & 0.131 (DEIRA) & 0.120 & 0.128   & -0.14 \\
  [8pt] \hline
   \multicolumn{5}{c}{$u_p=4.0$, $\Delta u_{fs}=1.063$} \\
  \hline
  $\beta_1$     & -- & 2.031  & 2.117  & -- \\
  $\beta_2$     & -- & 0.2879 & 0.2812 & -- \\
  $\delta_{pb}$ & -- & 0.4740 & 0.4800 & -- \\
  $\sigma_{sp}$ & 8.219 (EOS) & 8.147 & 10.75   & 12.45 \\
  $\bar{m}_{sp}$ & 0.568 (DEIRA) & 0.477 & 0.492  & 0.277 \\
\end{tabular}
\end{ruledtabular}
\end{table}

The relative impact of the pu-nonlinearity term depends on the value of $\beta_2$, which is insensitive to $u_p$ and reaches its maximum for the maximum possible spall strength. But even then the value $\beta_2 \approx 0.3$, obtained in our simulations and quoted in Table~\ref{t:4}, appears as rather an absolute upper limit; more typical for liquid metals at maximum possible tension would be $\beta_2 \lesssim 0.1$--0.15, like the value $\beta_2\approx 0.12$ obtained in the MD simulations for liquid copper \cite[]{Cai_Wu.2017}. The latter implies that the pu-nonlinearity should be taken into account near the theoretical limit for $\sigma_{th}$, but can be ignored in experiments where the spall failure occurs at tensions 3--10 times below this limit.

A characteristic feature of the formula (\ref{ss:sig_sp=}) is that the contributions of its three PA correction terms tend to compensate for one another. For large $u_p$, the decreasing impedance $\rho_{01}c_{01}$ of the zero-pressure release state strongly suppresses the impact of the $1+\delta_{pb}$ factor; for $\beta_2 \approx 0.2$--0.25 the effect of the pb-attenuation is practically canceled by that of the pu-nonlinearity term. The latter explains why it is not easy to establish a clear tendency in the role of nonlinearity corrections \cite[]{Cai_Wu.2017}, and why in practice the simple formula (\ref{PA:sig_ac=}) may frequently be more accurate than more complex expressions with only a partial account for the nonlinearity effects. On the whole, the data in Table~\ref{t:4} confirm that Eqs.~(\ref{ss:sig_sp=}) and (\ref{ss:m_sp=}) provide a significant improvement over the traditional acoustic formulae for interpretation of spall experiments with symmetric initial loading.

Our final remark is on the applicability of the acoustic limit. Naively, one might expect that in the limit of small amplitudes of both the loading shock and the spall pulse, i.e.\ in the limit of $\beta_1,\beta_2 \to 0$, $z_0\to 1$, the acoustic formula (\ref{PA:sig_ac=}) should be asymptotically exact. Our present analysis demonstrates that, whenever a pull-back shock is present, this is not the case. Instead, for symmetrically loaded impact plates, the correct small-amplitude limit is given by the formula
\begin{equation}\label{ss:sig_neo-ac=}
  \sigma_{sp} = \frac{1}{2} (1+\delta_{00})\rho_0 c_0 \Delta u_{fs}
  = \frac{1}{\sqrt{2}} \,\rho_0 c_0 \Delta u_{fs},
\end{equation}
which exceeds the acoustic result by 41\%, and which would be suitable for interpretation of experiments where $\beta_1 \lesssim 0.3$ and $\beta_2 \lesssim 0.05$.

\section{Conclusions \label{s:ccl}}

The paper presents the results of detailed analysis of the phase-flip (PF) model of spallation in pure liquids, where spall failure is treated as an infinitely fast irreversible liquid-gas phase transition. The spall strength is set equal to its theoretical maximum determined by the thermodynamic stability limit, i.e.\ by the negative pressure on the liquid branch of the spinodal in a two-phase EOS. The main advantage of the proposed model is that it can be easily and in a thermodynamically fully consistent way incorporated into the framework of ideal hydrodynamics. Unlike in the approach proposed in Ref.~\onlinecite{Grady1988,Grady1996}, energy dissipation due to spall fracture  is described without invoking the concepts of surface tension and/or viscosity, simply as the entropy jumps in the PF transitions and in shocks generated by the PF pressure jumps.

The process of spall fracture in the proposed model develops as a sudden appearance of one or more ``fog'' bubbles behind hypersonic spall fronts (representing a new type of discontinuity in ideal hydrodynamics) that are launched from one or more isolated inception points at local pressure minima. In our context, the term ``fog'' refers to the equilibrium matter state, to which a metastable liquid relaxes in the process of spinodal decomposition. The volume and mass of each fog bubble continue to grow until the spall front, precipitating the PF transition, is overtaken by the pull-back compression shock.

To illustrate how the PF model could be used in practice, the results of numerical modeling of a symmetric plate impact experiment are presented. In most aspects, they look quite similar to those found in the MD simulation of an analogous problem \cite[]{Cai_Wu.2017} --- but obtained at immeasurably lower computational cost. The principal difference between the two approaches manifests itself only in the microstructure of the fracture zone: while in the MD case it is stipulated by the statistics and the inherent atomic scale, in the fully deterministic PF approach the sizes and distribution of liquid fragments are determined by small flow perturbations and inhomogeneities of the initial state. When equipped with an adequate two-phase EOS, the PF model has a potential to become a powerful tool for modeling experiments where spallation and cavitation in liquids is an issue.

Using the PF numerical results as a reference solution, we derive new post-acoustic formulae for evaluation of the strain rate, the fractured mass fraction, and the spall strength in plate impact experiments where these quantities are to be inferred from the measured free-surface velocity. Though valid for the particular case of a symmetric plate impact, the new approximate formula (\ref{ss:sig_sp=}) for $\sigma_{th}$ reveals a non-trivial interplay of the three nonlinear correction terms, which in reality often compensate for one another. A notable new result is that the correction factor for the pb-shock attenuation never becomes close to unity, and in the limit $u_p\ll c_0$ of weak initial loads approaches a universal value $1+\delta_{00} =\sqrt{2}$ for any equation of state.

The key assumption of instant phase relaxation, on which the proposed PF model is based, is justified by explosive increase of the rate of homogeneous bubble nucleation near the limit of thermodynamic stability. In reality, the respective phase relaxation timescale $\tau_{pr}$ remains finite and limited to $\tau_{pr} \gtrsim 1$--3~ps, as is indicated by pressure relaxation dynamics in MD simulations. Hence, the hypersonic spall fronts, discussed in Sect.~\ref{s:spf}, have a finite width of $\sim u_{spf} \tau_{pr}$, where $u_{spf}$ is the spall-front velocity in the comoving system. Consequently, the proposed model becomes applicable on time scales $t \gg \tau_{pr}$ and length scales $l \gg c_0\tau_{pr} \gtrsim 10$~nm, where $c_0$ is the sound velocity in the relevant zero-pressure metastable state. Also, the length scale of emerging fracture zones must exceed the typical size of critical vapor bubbles, which again sets a limit of $l \gg 10$~nm. If the strain-rate dependence of the spall strength is to be accounted for, the present  assumption of the spinodal-limited spall strength can be relaxed along the lines proposed in Refs.~\onlinecite{Dekel_Eliezer.1998}, \cite{Pisarev_Kuksin.2009}, \cite{Faik_Basko.2012} and in other similar works.

\begin{acknowledgments}
This work was funded by the Keldysh Institute of Applied Mathematics of the Russian Academy of Sciences under the state contract with the Ministry of Science and Higher Education of the Russian Federation.
\end{acknowledgments}

\appendix

\section{Derivation of formula (\ref{ss:del_pb=}) for the pb-attenuation correction \label{s:A}}

We begin by ascertaining that the trajectory $\bar{m}_{pb}(\bar{t})$ of the pb shock, as it propagates from point $M_{sp}$ to point $M_2$ in Fig.~\ref{f:10} against the background of the R2 simple wave where all flow characteristics are functions of a single parameter $u'<0$ [see Eqs.~(\ref{isPA:p=})--(\ref{isPA:c=})], is governed by the equation
\begin{equation}\label{A:dm_pb/dt=}
  \frac{d\bar{m}_{pb}}{d\bar{t}} \equiv \zeta(\bar{u}') =1-b|\bar{u}'|.
\end{equation}
Indeed, the trajectory $m(t)$ of any shock front, traveling with a speed $D_{if}$ relative to the fluid in an initial state $(v_i,p_i)$, obeys the equation \cite[\S~16, Ch.~I]{Zeldovich_Raizer2012}
\begin{equation}\label{A:dm_sh/dt=}
  \frac{dm}{dt} =\rho_i D_{if} = \left(\frac{p_f-p_i}{v_i-v_f}\right)^{1/2},
\end{equation}
where $(v_f,p_f)$ is the final post-shock state. Important here is that the shock impedance (\ref{A:dm_sh/dt=}) remains invariant by interchanging the initial and final states. In our context, where the compression Hugoniot for a final state $p_f=p_H(u')>0$ at $u'>0$ and the initial state $p_i=p_H(0)=0$ is given by Eqs.~(\ref{isPA:p_H=}), (\ref{isPA:rho_H=}), the normalized shock impedance takes the form
\begin{equation}\label{A:z_sh=}
  \frac{d\bar{m}}{d\bar{t}} =1+b\bar{u}'.
\end{equation}
If we consider now the pb compression shock, whose initial pressure $p_i=p_H(u')<0$ is given by Eq.~(\ref{isPA:p_H=}) [identical to Eq.~(\ref{isPA:p=})] for some $u'<0$ and the final pressure $p_f=p_H(0)=0$, we conclude that, due to the $i\rightleftarrows f$ interchange symmetry, its impedance must be equal to that of the \emph{rarefaction shock} (perhaps non-physical but mathematically conceivable) represented by the extension of Eqs.~(\ref{isPA:p_H=}), (\ref{isPA:rho_H=}), and (\ref{A:z_sh=}) into the $u'<0$ domain, i.e.\ by Eq.~(\ref{A:dm_pb/dt=}). Note that, unlike its normalized impedance $\zeta$, the front speed of the pb shock
\begin{equation}\label{A:D_pb=}
  D_{pb} =c_{01} -(b-1)|u'|
\end{equation}
is not equal to the extension of Eq.~(\ref{isPA:D=}) to negative $u'$. The fact that $\rho_H(u')$ in Eq.~(\ref{isPA:rho_H=}) is not exactly equal to $\rho(\bar{u}')$ from Eq.~(\ref{isPA:rho=}) has no practical significance because, when expanded in powers of $\bar{u}'$, the two formulae begin to diverge in $O(\bar{u}'^3)$ terms only, which can be safely ignored for $|\bar{u}'| \leq |\bar{u}'_{sp}| \lesssim 0.1$--0.15 encountered in reality. Of negligible effect is also the essentially positive pressure behind the pb shock, which, being limited from above by $p_{EQ,sp}$ (see Sect.~\ref{s:spf}), can practically always be neglected in comparison with $\sigma_{sp}$, as is seen from Table~\ref{t:2}.

Next, we introduce an auxiliary characteristic $C^+_{o2}$ into the chart in Fig.~\ref{f:10}, which belongs to the $C^+$ fan of the R2 wave and crosses the right-reflected characteristic $C^-_{h3}$ at the same point $M_2$ as the pb shock. Because for $u'<0$ the acoustic impedance (\ref{isPA:z=}) is smaller than the pb-shock impedance $\zeta$, $C^+_{o2}$ is straight and lies below the pb-shock trajectory everywhere between $M_2$ and its intersection with the $C^-_{t1}$ characteristic at point $M_{o2}$ with coordinates $(\bar{m}_{o2},\bar{t}_{o2})$, which lie in the range
\begin{equation}\label{A:m_o2<=}
  \bar{m}_{sp} < \bar{m}_{o2}< \bar{m}_{1m}, \quad \bar{t}_{1m}< \bar{t}_{o2}< \bar{t}_{sp}.
\end{equation}
When applied to points $M_{o2}$ and $M_{sp}$, the LU approximation (\ref{sr:LU-app=}) takes the form
\begin{equation}\label{A:LU=}
  \frac{u_1}{m_{1m}} = \frac{u_1-|u'_2|}{m_{o2}} = \frac{u_1-|u'_{sp}|}{m_{sp}},
\end{equation}
which relates $\bar{m}_{o2}$ and $\bar{m}_{sp}$ to the known quantity $\bar{m}_{1m}$, leading, in particular, to Eq.~(\ref{ss:m_sp=}).

Now, the two unknown quantities $\delta_{pb}$ and $\bar{m}_{2m}$ can be calculated from the system of two equations
\begin{equation}\label{A:Eq_1=}
  1-\bar{m}_{1m} +\frac{1-\bar{m}_{2m}}{(1+z_2)/2} - \frac{\bar{m}_{2m}-\bar{m}_{o2}}{z_2}
  -\frac{\bar{m}_{1m}-\bar{m}_{o2}}{(1+z_2)/2} = 0,
\end{equation}
\begin{equation} \label{A:Eq_2=}
  \frac{\bar{m}_{2m}-\bar{m}_{o2}}{z_2} -
  \frac{\bar{m}_{2m}-\bar{m}_{sp}}{(\zeta_2+\zeta_{sp})/2} -
  \frac{\bar{m}_{o2}-\bar{m}_{sp}}{(z_2+z_{sp})/2} = 0,
\end{equation}
obtained, respectively, by applying the $\oint d\bar{t}=\oint z^{-1}\, d\bar{t}=0$ circulation rule to the two closed contours $M_1$-$\bar{t}_2$-$M_2$-$M_{o2}$-$M_1$ and $M_{o2}$-$M_2$-$M_{sp}$-$M_{o2}$ in the $(\bar{m},\bar{t})$ plane, and by using the mean impedance values along the curvilinear characteristic segments and the pb-shock trajectory. Here
\begin{equation}\label{A:z2,zeta_2=}
  z_2=1-2\beta_2, \quad \zeta_2=1-\beta_2,
\end{equation}
are the normalized acoustic and pb-shock impedances at point $M_2$, and
\begin{equation}\label{A:zeta_2=}
  z_{sp}= 1-2\beta_2(1+\delta_{pb}), \quad \zeta_{sp}= 1-\beta_2(1+\delta_{pb})
\end{equation}
are the same impedances at point $M_{sp}$; $\beta_2$ is defined in Eq.~(\ref{ss:bet_2=}). Equations (\ref{A:Eq_1=}) and (\ref{A:Eq_2=}) are easily transformed to
\begin{eqnarray} \label{A:Eq_1a=}
  \bar{m}_{2m}-\bar{m}_{o2} &=& (1-\bar{m}_{1m}) \frac{z_2(3+z_2)}{1+3z_2},
  \\ \label{A:Eq_2a=}
  \frac{\bar{m}_{2m}-\bar{m}_{o2}}{\bar{m}_{o2}-\bar{m}_{sp}} &=&
  \frac{2z_2(z_2+z_{sp}+\zeta_2+\zeta_{sp})}{(z_2+z_{sp})(\zeta_2+
  \zeta_{sp}-2z_2)},
\end{eqnarray}
while from Eq.~(\ref{A:LU=}) we get
\begin{equation}\label{A:Eq_3a=}
  \bar{m}_{o2}-\bar{m}_{sp}= \bar{m}_{1m} (\beta_2/\beta_1) \delta_{pb}.
\end{equation}
Having substituted Eqs.~(\ref{A:z2,zeta_2=})--(\ref{A:Eq_1a=}) together with Eqs.~(\ref{A:Eq_3a=}) and (\ref{sr:bm_1m_PA=}) into Eq.~(\ref{A:Eq_2a=}), we obtain the quadratic equation
\begin{equation}\label{A:Eq-del_pb=}
  (3\mu+1)\beta_2 \delta_{pb}^2 -2\omega \delta_{pb} +2(1-2\beta_2)=0
\end{equation}
for the attenuation correction $\delta_{pb}$, where $\mu$ and $\omega$ are defined in Eq.~(\ref{ss:mu,omega=}). The basic equations (\ref{A:Eq_1=}), (\ref{A:Eq_2=}) and the ensuing formulae are physically meaningful only if the impedance $1\geq z_2\geq 0$, i.e.\ under the condition
\begin{equation}\label{A:bet_2<=}
  0 \leq \beta_2 \leq \frac{1}{2}.
\end{equation}
It is not difficult to prove that the above condition guarantees positiveness of the discriminant in Eq.~(\ref{A:Eq-del_pb=}).

\bibliography{Basko_pf-cav20_Arxiv}

\begin{thebibliography}{43}%
\makeatletter
\providecommand \@ifxundefined [1]{%
 \@ifx{#1\undefined}
}%
\providecommand \@ifnum [1]{%
 \ifnum #1\expandafter \@firstoftwo
 \else \expandafter \@secondoftwo
 \fi
}%
\providecommand \@ifx [1]{%
 \ifx #1\expandafter \@firstoftwo
 \else \expandafter \@secondoftwo
 \fi
}%
\providecommand \natexlab [1]{#1}%
\providecommand \enquote  [1]{``#1''}%
\providecommand \bibnamefont  [1]{#1}%
\providecommand \bibfnamefont [1]{#1}%
\providecommand \citenamefont [1]{#1}%
\providecommand \href@noop [0]{\@secondoftwo}%
\providecommand \href [0]{\begingroup \@sanitize@url \@href}%
\providecommand \@href[1]{\@@startlink{#1}\@@href}%
\providecommand \@@href[1]{\endgroup#1\@@endlink}%
\providecommand \@sanitize@url [0]{\catcode `\\12\catcode `\$12\catcode
  `\&12\catcode `\#12\catcode `\^12\catcode `\_12\catcode `\%12\relax}%
\providecommand \@@startlink[1]{}%
\providecommand \@@endlink[0]{}%
\providecommand \url  [0]{\begingroup\@sanitize@url \@url }%
\providecommand \@url [1]{\endgroup\@href {#1}{\urlprefix }}%
\providecommand \urlprefix  [0]{URL }%
\providecommand \Eprint [0]{\href }%
\providecommand \doibase [0]{http://dx.doi.org/}%
\providecommand \selectlanguage [0]{\@gobble}%
\providecommand \bibinfo  [0]{\@secondoftwo}%
\providecommand \bibfield  [0]{\@secondoftwo}%
\providecommand \translation [1]{[#1]}%
\providecommand \BibitemOpen [0]{}%
\providecommand \bibitemStop [0]{}%
\providecommand \bibitemNoStop [0]{.\EOS\space}%
\providecommand \EOS [0]{\spacefactor3000\relax}%
\providecommand \BibitemShut  [1]{\csname bibitem#1\endcsname}%
\let\auto@bib@innerbib\@empty
\bibitem [{\citenamefont {Antoun}\ \emph {et~al.}(2003)\citenamefont {Antoun},
  \citenamefont {Seaman}, \citenamefont {Curran}, \citenamefont {Kanel},
  \citenamefont {Razorenov},\ and\ \citenamefont {Utkin}}]{Antoun_Seaman.2003}%
  \BibitemOpen
  \bibfield  {author} {\bibinfo {author} {\bibfnamefont {T.}~\bibnamefont
  {Antoun}}, \bibinfo {author} {\bibfnamefont {L.}~\bibnamefont {Seaman}},
  \bibinfo {author} {\bibfnamefont {D.~R.}\ \bibnamefont {Curran}}, \bibinfo
  {author} {\bibfnamefont {G.~I.}\ \bibnamefont {Kanel}}, \bibinfo {author}
  {\bibfnamefont {S.~V.}\ \bibnamefont {Razorenov}}, \ and\ \bibinfo {author}
  {\bibfnamefont {A.~V.}\ \bibnamefont {Utkin}},\ }\href@noop {} {\emph
  {\bibinfo {title} {Spall {F}racture}}},\ Shock Wave and High Pressure
  Phenomena\ (\bibinfo  {publisher} {Springer-Verlag New York},\ \bibinfo
  {year} {2003})\BibitemShut {NoStop}%
\bibitem [{\citenamefont {Kanel}, \citenamefont {Fortov},\ and\ \citenamefont
  {Razorenov}(2007)}]{Kanel_Fortov.2007}%
  \BibitemOpen
  \bibfield  {author} {\bibinfo {author} {\bibfnamefont {G.~I.}\ \bibnamefont
  {Kanel}}, \bibinfo {author} {\bibfnamefont {V.~E.}\ \bibnamefont {Fortov}}, \
  and\ \bibinfo {author} {\bibfnamefont {S.~V.}\ \bibnamefont {Razorenov}},\
  }\bibfield  {title} {\enquote {\bibinfo {title} {Shock waves in
  condensed-state physics},}\ }\href {\doibase 10.1070/pu2007v050n08abeh006327}
  {\bibfield  {journal} {\bibinfo  {journal} {Physics-Uspekhi}\ }\textbf
  {\bibinfo {volume} {50}},\ \bibinfo {pages} {771--791} (\bibinfo {year}
  {2007})}\BibitemShut {NoStop}%
\bibitem [{\citenamefont {Kanel}\ \emph {et~al.}(2015)\citenamefont {Kanel},
  \citenamefont {Savinykh}, \citenamefont {Garkushin},\ and\ \citenamefont
  {Razorenov}}]{Kanel_Savinykh.2015}%
  \BibitemOpen
  \bibfield  {author} {\bibinfo {author} {\bibfnamefont {G.~I.}\ \bibnamefont
  {Kanel}}, \bibinfo {author} {\bibfnamefont {A.~S.}\ \bibnamefont {Savinykh}},
  \bibinfo {author} {\bibfnamefont {G.~V.}\ \bibnamefont {Garkushin}}, \ and\
  \bibinfo {author} {\bibfnamefont {S.~V.}\ \bibnamefont {Razorenov}},\
  }\bibfield  {title} {\enquote {\bibinfo {title} {Dynamic strength of tin and
  lead melts},}\ }\href {\doibase 10.1134/S0021364015200059} {\bibfield
  {journal} {\bibinfo  {journal} {JETP Letters}\ }\textbf {\bibinfo {volume}
  {102}},\ \bibinfo {pages} {548--551} (\bibinfo {year} {2015})}\BibitemShut
  {NoStop}%
\bibitem [{\citenamefont {Zaretsky}(2016)}]{Zaretsky2016}%
  \BibitemOpen
  \bibfield  {author} {\bibinfo {author} {\bibfnamefont {E.~B.}\ \bibnamefont
  {Zaretsky}},\ }\bibfield  {title} {\enquote {\bibinfo {title} {Experimental
  determination of the dynamic tensile strength of liquid {S}n, {P}b, and
  {Z}n},}\ }\href {\doibase 10.1063/1.4958798} {\bibfield  {journal} {\bibinfo
  {journal} {Journal of Applied Physics}\ }\textbf {\bibinfo {volume} {120}},\
  \bibinfo {pages} {025902} (\bibinfo {year} {2016})}\BibitemShut {NoStop}%
\bibitem [{\citenamefont {de~Ress\'eguier}\ \emph {et~al.}(2007)\citenamefont
  {de~Ress\'eguier}, \citenamefont {Signor}, \citenamefont {Dragon},
  \citenamefont {Boustie}, \citenamefont {Roy},\ and\ \citenamefont
  {Llorca}}]{Resseguier_Signor.2007}%
  \BibitemOpen
  \bibfield  {author} {\bibinfo {author} {\bibfnamefont {T.}~\bibnamefont
  {de~Ress\'eguier}}, \bibinfo {author} {\bibfnamefont {L.}~\bibnamefont
  {Signor}}, \bibinfo {author} {\bibfnamefont {A.}~\bibnamefont {Dragon}},
  \bibinfo {author} {\bibfnamefont {M.}~\bibnamefont {Boustie}}, \bibinfo
  {author} {\bibfnamefont {G.}~\bibnamefont {Roy}}, \ and\ \bibinfo {author}
  {\bibfnamefont {F.}~\bibnamefont {Llorca}},\ }\bibfield  {title} {\enquote
  {\bibinfo {title} {Experimental investigation of liquid spall in laser
  shock-loaded tin},}\ }\href {\doibase 10.1063/1.2400800} {\bibfield
  {journal} {\bibinfo  {journal} {Journal of Applied Physics}\ }\textbf
  {\bibinfo {volume} {101}},\ \bibinfo {pages} {013506} (\bibinfo {year}
  {2007})}\BibitemShut {NoStop}%
\bibitem [{\citenamefont {Agranat}\ \emph {et~al.}(2010)\citenamefont
  {Agranat}, \citenamefont {Anisimov}, \citenamefont {Ashitkov}, \citenamefont
  {Zhakhovskii}, \citenamefont {Inogamov}, \citenamefont {Komarov},
  \citenamefont {Ovchinnikov}, \citenamefont {Fortov}, \citenamefont
  {Khokhlov},\ and\ \citenamefont {Shepelev}}]{Agranat_Anisimov.2010}%
  \BibitemOpen
  \bibfield  {author} {\bibinfo {author} {\bibfnamefont {M.~B.}\ \bibnamefont
  {Agranat}}, \bibinfo {author} {\bibfnamefont {S.~I.}\ \bibnamefont
  {Anisimov}}, \bibinfo {author} {\bibfnamefont {S.~I.}\ \bibnamefont
  {Ashitkov}}, \bibinfo {author} {\bibfnamefont {V.~V.}\ \bibnamefont
  {Zhakhovskii}}, \bibinfo {author} {\bibfnamefont {N.~A.}\ \bibnamefont
  {Inogamov}}, \bibinfo {author} {\bibfnamefont {P.~S.}\ \bibnamefont
  {Komarov}}, \bibinfo {author} {\bibfnamefont {A.~V.}\ \bibnamefont
  {Ovchinnikov}}, \bibinfo {author} {\bibfnamefont {V.~E.}\ \bibnamefont
  {Fortov}}, \bibinfo {author} {\bibfnamefont {V.~A.}\ \bibnamefont
  {Khokhlov}}, \ and\ \bibinfo {author} {\bibfnamefont {V.~V.}\ \bibnamefont
  {Shepelev}},\ }\bibfield  {title} {\enquote {\bibinfo {title} {Strength
  properties of an aluminum melt at extremely high tension rates under the
  action of femtosecond laser pulses},}\ }\href {\doibase
  10.1134/S0021364010090080} {\bibfield  {journal} {\bibinfo  {journal} {JETP
  Letters}\ }\textbf {\bibinfo {volume} {91}},\ \bibinfo {pages} {471--477}
  (\bibinfo {year} {2010})}\BibitemShut {NoStop}%
\bibitem [{\citenamefont {Krivokorytov}\ \emph {et~al.}(2017)\citenamefont
  {Krivokorytov}, \citenamefont {Vinokhodov}, \citenamefont {Sidelnikov},
  \citenamefont {Krivtsun}, \citenamefont {Kompanets}, \citenamefont {Lash},
  \citenamefont {Koshelev},\ and\ \citenamefont
  {Medvedev}}]{Krivokorytov_Vinokhodov.2017}%
  \BibitemOpen
  \bibfield  {author} {\bibinfo {author} {\bibfnamefont {M.~S.}\ \bibnamefont
  {Krivokorytov}}, \bibinfo {author} {\bibfnamefont {A.~Y.}\ \bibnamefont
  {Vinokhodov}}, \bibinfo {author} {\bibfnamefont {Y.~V.}\ \bibnamefont
  {Sidelnikov}}, \bibinfo {author} {\bibfnamefont {V.~M.}\ \bibnamefont
  {Krivtsun}}, \bibinfo {author} {\bibfnamefont {V.~O.}\ \bibnamefont
  {Kompanets}}, \bibinfo {author} {\bibfnamefont {A.~A.}\ \bibnamefont {Lash}},
  \bibinfo {author} {\bibfnamefont {K.~N.}\ \bibnamefont {Koshelev}}, \ and\
  \bibinfo {author} {\bibfnamefont {V.~V.}\ \bibnamefont {Medvedev}},\
  }\bibfield  {title} {\enquote {\bibinfo {title} {Cavitation and spallation in
  liquid metal droplets produced by subpicosecond pulsed laser radiation},}\
  }\href {\doibase 10.1103/PhysRevE.95.031101} {\bibfield  {journal} {\bibinfo
  {journal} {Phys. Rev. E}\ }\textbf {\bibinfo {volume} {95}},\ \bibinfo
  {pages} {031101} (\bibinfo {year} {2017})}\BibitemShut {NoStop}%
\bibitem [{\citenamefont {Stan}\ \emph {et~al.}(2016)\citenamefont {Stan},
  \citenamefont {Willmott}, \citenamefont {Stone}, \citenamefont {Koglin},
  \citenamefont {Liang}, \citenamefont {Aquila}, \citenamefont {Robinson},
  \citenamefont {Gumerlock}, \citenamefont {Blaj}, \citenamefont {Sierra},
  \citenamefont {Boutet}, \citenamefont {Guillet}, \citenamefont {Curtis},
  \citenamefont {Vetter}, \citenamefont {Loos}, \citenamefont {Turner},\ and\
  \citenamefont {Decker}}]{Stan_Willmott.2016}%
  \BibitemOpen
  \bibfield  {author} {\bibinfo {author} {\bibfnamefont {C.~A.}\ \bibnamefont
  {Stan}}, \bibinfo {author} {\bibfnamefont {P.~R.}\ \bibnamefont {Willmott}},
  \bibinfo {author} {\bibfnamefont {H.~A.}\ \bibnamefont {Stone}}, \bibinfo
  {author} {\bibfnamefont {J.~E.}\ \bibnamefont {Koglin}}, \bibinfo {author}
  {\bibfnamefont {M.}~\bibnamefont {Liang}}, \bibinfo {author} {\bibfnamefont
  {A.~L.}\ \bibnamefont {Aquila}}, \bibinfo {author} {\bibfnamefont {J.~S.}\
  \bibnamefont {Robinson}}, \bibinfo {author} {\bibfnamefont {K.~L.}\
  \bibnamefont {Gumerlock}}, \bibinfo {author} {\bibfnamefont {G.}~\bibnamefont
  {Blaj}}, \bibinfo {author} {\bibfnamefont {R.~G.}\ \bibnamefont {Sierra}},
  \bibinfo {author} {\bibfnamefont {S.}~\bibnamefont {Boutet}}, \bibinfo
  {author} {\bibfnamefont {S.~A.~H.}\ \bibnamefont {Guillet}}, \bibinfo
  {author} {\bibfnamefont {R.~H.}\ \bibnamefont {Curtis}}, \bibinfo {author}
  {\bibfnamefont {S.~L.}\ \bibnamefont {Vetter}}, \bibinfo {author}
  {\bibfnamefont {H.}~\bibnamefont {Loos}}, \bibinfo {author} {\bibfnamefont
  {J.~L.}\ \bibnamefont {Turner}}, \ and\ \bibinfo {author} {\bibfnamefont
  {F.-J.}\ \bibnamefont {Decker}},\ }\bibfield  {title} {\enquote {\bibinfo
  {title} {Negative pressures and spallation in water drops subjected to
  nanosecond shock waves},}\ }\href {\doibase 10.1021/acs.jpclett.6b00687}
  {\bibfield  {journal} {\bibinfo  {journal} {J. Phys. Chem. Lett.}\ }\textbf
  {\bibinfo {volume} {7}},\ \bibinfo {pages} {2055--2062} (\bibinfo {year}
  {2016})}\BibitemShut {NoStop}%
\bibitem [{\citenamefont {Basko}\ \emph {et~al.}(2017)\citenamefont {Basko},
  \citenamefont {Krivokorytov}, \citenamefont {Vinokhodov}, \citenamefont
  {Sidelnikov}, \citenamefont {Krivtsun}, \citenamefont {Medvedev},
  \citenamefont {Kim}, \citenamefont {Kompanets}, \citenamefont {Lash},\ and\
  \citenamefont {Koshelev}}]{Basko_Krivokor.2017}%
  \BibitemOpen
  \bibfield  {author} {\bibinfo {author} {\bibfnamefont {M.~M.}\ \bibnamefont
  {Basko}}, \bibinfo {author} {\bibfnamefont {M.~S.}\ \bibnamefont
  {Krivokorytov}}, \bibinfo {author} {\bibfnamefont {A.~Y.}\ \bibnamefont
  {Vinokhodov}}, \bibinfo {author} {\bibfnamefont {Y.~V.}\ \bibnamefont
  {Sidelnikov}}, \bibinfo {author} {\bibfnamefont {V.~M.}\ \bibnamefont
  {Krivtsun}}, \bibinfo {author} {\bibfnamefont {V.~V.}\ \bibnamefont
  {Medvedev}}, \bibinfo {author} {\bibfnamefont {D.~A.}\ \bibnamefont {Kim}},
  \bibinfo {author} {\bibfnamefont {V.~O.}\ \bibnamefont {Kompanets}}, \bibinfo
  {author} {\bibfnamefont {A.~A.}\ \bibnamefont {Lash}}, \ and\ \bibinfo
  {author} {\bibfnamefont {K.~N.}\ \bibnamefont {Koshelev}},\ }\bibfield
  {title} {\enquote {\bibinfo {title} {Fragmentation dynamics of {liquid-metal}
  droplets under ultra-short laser pulses},}\ }\href {\doibase
  10.1088/1612-202X/aa539b} {\bibfield  {journal} {\bibinfo  {journal} {Laser
  Physics Letters}\ }\textbf {\bibinfo {volume} {14}},\ \bibinfo {pages}
  {036001} (\bibinfo {year} {2017})}\BibitemShut {NoStop}%
\bibitem [{\citenamefont {Grigoryev}\ \emph {et~al.}(2018)\citenamefont
  {Grigoryev}, \citenamefont {Lakatosh}, \citenamefont {Krivokorytov},
  \citenamefont {Zhakhovsky}, \citenamefont {Dyachkov}, \citenamefont
  {Ilnitsky}, \citenamefont {Migdal}, \citenamefont {Inogamov}, \citenamefont
  {Vinokhodov}, \citenamefont {Kompanets}, \citenamefont {Sidelnikov},
  \citenamefont {Krivtsun}, \citenamefont {Koshelev},\ and\ \citenamefont
  {Medvedev}}]{Grigoryev_Lakatosh.2018}%
  \BibitemOpen
  \bibfield  {author} {\bibinfo {author} {\bibfnamefont {S.~Y.}\ \bibnamefont
  {Grigoryev}}, \bibinfo {author} {\bibfnamefont {B.~V.}\ \bibnamefont
  {Lakatosh}}, \bibinfo {author} {\bibfnamefont {M.~S.}\ \bibnamefont
  {Krivokorytov}}, \bibinfo {author} {\bibfnamefont {V.~V.}\ \bibnamefont
  {Zhakhovsky}}, \bibinfo {author} {\bibfnamefont {S.~A.}\ \bibnamefont
  {Dyachkov}}, \bibinfo {author} {\bibfnamefont {D.~K.}\ \bibnamefont
  {Ilnitsky}}, \bibinfo {author} {\bibfnamefont {K.~P.}\ \bibnamefont
  {Migdal}}, \bibinfo {author} {\bibfnamefont {N.~A.}\ \bibnamefont
  {Inogamov}}, \bibinfo {author} {\bibfnamefont {A.~Y.}\ \bibnamefont
  {Vinokhodov}}, \bibinfo {author} {\bibfnamefont {V.~O.}\ \bibnamefont
  {Kompanets}}, \bibinfo {author} {\bibfnamefont {Y.~V.}\ \bibnamefont
  {Sidelnikov}}, \bibinfo {author} {\bibfnamefont {V.~M.}\ \bibnamefont
  {Krivtsun}}, \bibinfo {author} {\bibfnamefont {K.~N.}\ \bibnamefont
  {Koshelev}}, \ and\ \bibinfo {author} {\bibfnamefont {V.~V.}\ \bibnamefont
  {Medvedev}},\ }\bibfield  {title} {\enquote {\bibinfo {title} {Expansion and
  fragmentation of a liquid-metal droplet by a short laser pulse},}\ }\href
  {\doibase 10.1103/PhysRevApplied.10.064009} {\bibfield  {journal} {\bibinfo
  {journal} {Phys. Rev. Applied}\ }\textbf {\bibinfo {volume} {10}},\ \bibinfo
  {pages} {064009} (\bibinfo {year} {2018})}\BibitemShut {NoStop}%
\bibitem [{\citenamefont {Pisarev}\ \emph {et~al.}(2009)\citenamefont
  {Pisarev}, \citenamefont {Kuksin}, \citenamefont {Norman}, \citenamefont
  {Stegailov},\ and\ \citenamefont {Yanilkin}}]{Pisarev_Kuksin.2009}%
  \BibitemOpen
  \bibfield  {author} {\bibinfo {author} {\bibfnamefont {V.~V.}\ \bibnamefont
  {Pisarev}}, \bibinfo {author} {\bibfnamefont {A.~Y.}\ \bibnamefont {Kuksin}},
  \bibinfo {author} {\bibfnamefont {G.~E.}\ \bibnamefont {Norman}}, \bibinfo
  {author} {\bibfnamefont {V.~V.}\ \bibnamefont {Stegailov}}, \ and\ \bibinfo
  {author} {\bibfnamefont {A.~V.}\ \bibnamefont {Yanilkin}},\ }\bibfield
  {title} {\enquote {\bibinfo {title} {Microscopic theory and kinetic model of
  spall in liquids},}\ }\href {\doibase 10.1063/1.3295263} {\bibfield
  {journal} {\bibinfo  {journal} {AIP Conference Proceedings}\ }\textbf
  {\bibinfo {volume} {1195}},\ \bibinfo {pages} {801--804} (\bibinfo {year}
  {2009})}\BibitemShut {NoStop}%
\bibitem [{\citenamefont {Cai}, \citenamefont {Wu},\ and\ \citenamefont
  {Luo}(2017)}]{Cai_Wu.2017}%
  \BibitemOpen
  \bibfield  {author} {\bibinfo {author} {\bibfnamefont {Y.}~\bibnamefont
  {Cai}}, \bibinfo {author} {\bibfnamefont {H.~A.}\ \bibnamefont {Wu}}, \ and\
  \bibinfo {author} {\bibfnamefont {S.~N.}\ \bibnamefont {Luo}},\ }\bibfield
  {title} {\enquote {\bibinfo {title} {Spall strength of liquid copper and
  accuracy of the acoustic method},}\ }\href {\doibase 10.1063/1.4978251}
  {\bibfield  {journal} {\bibinfo  {journal} {Journal of Applied Physics}\
  }\textbf {\bibinfo {volume} {121}},\ \bibinfo {pages} {105901} (\bibinfo
  {year} {2017})}\BibitemShut {NoStop}%
\bibitem [{\citenamefont {Mayer}\ and\ \citenamefont
  {Mayer}(2020)}]{Mayer_Mayer2020}%
  \BibitemOpen
  \bibfield  {author} {\bibinfo {author} {\bibfnamefont {A.~E.}\ \bibnamefont
  {Mayer}}\ and\ \bibinfo {author} {\bibfnamefont {P.~N.}\ \bibnamefont
  {Mayer}},\ }\bibfield  {title} {\enquote {\bibinfo {title} {Strain rate
  dependence of spall strength for solid and molten lead and tin},}\ }\href
  {\doibase 10.1007/s10704-020-00440-8} {\bibfield  {journal} {\bibinfo
  {journal} {International Journal of Fracture}\ }\textbf {\bibinfo {volume}
  {222}},\ \bibinfo {pages} {171--195} (\bibinfo {year} {2020})}\BibitemShut
  {NoStop}%
\bibitem [{\citenamefont {Anisimov}\ \emph {et~al.}(1999)\citenamefont
  {Anisimov}, \citenamefont {Inogamov}, \citenamefont {Oparin}, \citenamefont
  {Rethfeld}, \citenamefont {Yabe}, \citenamefont {Ogawa},\ and\ \citenamefont
  {Fortov}}]{Anisimov_Inogamov.1999}%
  \BibitemOpen
  \bibfield  {author} {\bibinfo {author} {\bibfnamefont {S.~I.}\ \bibnamefont
  {Anisimov}}, \bibinfo {author} {\bibfnamefont {N.~A.}\ \bibnamefont
  {Inogamov}}, \bibinfo {author} {\bibfnamefont {A.~M.}\ \bibnamefont
  {Oparin}}, \bibinfo {author} {\bibfnamefont {B.}~\bibnamefont {Rethfeld}},
  \bibinfo {author} {\bibfnamefont {T.}~\bibnamefont {Yabe}}, \bibinfo {author}
  {\bibfnamefont {M.}~\bibnamefont {Ogawa}}, \ and\ \bibinfo {author}
  {\bibfnamefont {V.~E.}\ \bibnamefont {Fortov}},\ }\bibfield  {title}
  {\enquote {\bibinfo {title} {Pulsed laser evaporation: equation-of-state
  effects},}\ }\href {\doibase 10.1007/s003390051041} {\bibfield  {journal}
  {\bibinfo  {journal} {Applied Physics A}\ }\textbf {\bibinfo {volume} {69}},\
  \bibinfo {pages} {617--620} (\bibinfo {year} {1999})}\BibitemShut {NoStop}%
\bibitem [{\citenamefont {Colombier}\ \emph {et~al.}(2005)\citenamefont
  {Colombier}, \citenamefont {Combis}, \citenamefont {Bonneau}, \citenamefont
  {Harzic},\ and\ \citenamefont {Audouard}}]{Colombier_Combis.2005}%
  \BibitemOpen
  \bibfield  {author} {\bibinfo {author} {\bibfnamefont {J.~P.}\ \bibnamefont
  {Colombier}}, \bibinfo {author} {\bibfnamefont {P.}~\bibnamefont {Combis}},
  \bibinfo {author} {\bibfnamefont {F.}~\bibnamefont {Bonneau}}, \bibinfo
  {author} {\bibfnamefont {R.~L.}\ \bibnamefont {Harzic}}, \ and\ \bibinfo
  {author} {\bibfnamefont {E.}~\bibnamefont {Audouard}},\ }\bibfield  {title}
  {\enquote {\bibinfo {title} {Hydrodynamic simulations of metal ablation by
  femtosecond laser irradiation},}\ }\href {\doibase
  10.1103/PhysRevB.71.165406} {\bibfield  {journal} {\bibinfo  {journal}
  {Phys.\ Rev.\ B}\ }\textbf {\bibinfo {volume} {71}},\ \bibinfo {pages}
  {165406} (\bibinfo {year} {2005})}\BibitemShut {NoStop}%
\bibitem [{\citenamefont {Zhao}\ \emph {et~al.}(2011)\citenamefont {Zhao},
  \citenamefont {Mentrelli}, \citenamefont {Ruggeri},\ and\ \citenamefont
  {Sugiyama}}]{Zhao_Mentrelli.2011}%
  \BibitemOpen
  \bibfield  {author} {\bibinfo {author} {\bibfnamefont {N.}~\bibnamefont
  {Zhao}}, \bibinfo {author} {\bibfnamefont {A.}~\bibnamefont {Mentrelli}},
  \bibinfo {author} {\bibfnamefont {T.}~\bibnamefont {Ruggeri}}, \ and\
  \bibinfo {author} {\bibfnamefont {M.}~\bibnamefont {Sugiyama}},\ }\bibfield
  {title} {\enquote {\bibinfo {title} {Admissible shock waves and shock-induced
  phase transitions in a van der {W}aals fluid},}\ }\href {\doibase
  10.1063/1.3622772} {\bibfield  {journal} {\bibinfo  {journal} {Phys. Fluids}\
  }\textbf {\bibinfo {volume} {23}},\ \bibinfo {pages} {086101} (\bibinfo
  {year} {2011})}\BibitemShut {NoStop}%
\bibitem [{\citenamefont {Basko}(2018{\natexlab{a}})}]{Basko2018-PhF}%
  \BibitemOpen
  \bibfield  {author} {\bibinfo {author} {\bibfnamefont {M.~M.}\ \bibnamefont
  {Basko}},\ }\bibfield  {title} {\enquote {\bibinfo {title} {Centered
  rarefaction wave with a liquid-gas phase transition in the approximation of
  ``phase-flip'' hydrodynamics},}\ }\href {\doibase 10.1063/1.5064495}
  {\bibfield  {journal} {\bibinfo  {journal} {Physics of Fluids}\ }\textbf
  {\bibinfo {volume} {30}},\ \bibinfo {pages} {123306} (\bibinfo {year}
  {2018}{\natexlab{a}})}\BibitemShut {NoStop}%
\bibitem [{\citenamefont {Boteler}\ and\ \citenamefont
  {Sutherland}(2004)}]{Boteler_Sutherland2004}%
  \BibitemOpen
  \bibfield  {author} {\bibinfo {author} {\bibfnamefont {J.~M.}\ \bibnamefont
  {Boteler}}\ and\ \bibinfo {author} {\bibfnamefont {G.~T.}\ \bibnamefont
  {Sutherland}},\ }\bibfield  {title} {\enquote {\bibinfo {title} {Tensile
  failure of water due to shock wave interactions},}\ }\href {\doibase
  10.1063/1.1810635} {\bibfield  {journal} {\bibinfo  {journal} {Journal of
  Applied Physics}\ }\textbf {\bibinfo {volume} {96}},\ \bibinfo {pages}
  {6919--6924} (\bibinfo {year} {2004})}\BibitemShut {NoStop}%
\bibitem [{\citenamefont {Grady}(1988)}]{Grady1988}%
  \BibitemOpen
  \bibfield  {author} {\bibinfo {author} {\bibfnamefont {D.~E.}\ \bibnamefont
  {Grady}},\ }\bibfield  {title} {\enquote {\bibinfo {title} {The spall
  strength of condensed matter},}\ }\href {\doibase
  10.1016/0022-5096(88)90015-4} {\bibfield  {journal} {\bibinfo  {journal}
  {Journal of the Mechanics and Physics of Solids}\ }\textbf {\bibinfo {volume}
  {36}},\ \bibinfo {pages} {353--384} (\bibinfo {year} {1988})}\BibitemShut
  {NoStop}%
\bibitem [{\citenamefont {Grady}(1996)}]{Grady1996}%
  \BibitemOpen
  \bibfield  {author} {\bibinfo {author} {\bibfnamefont {D.~E.}\ \bibnamefont
  {Grady}},\ }\bibfield  {title} {\enquote {\bibinfo {title} {Spall and
  fragmentation in high-temperature metals},}\ }in\ \href {\doibase
  10.1007/978-1-4612-2320-7} {\emph {\bibinfo {booktitle} {L. Davison and D.E.
  Grady and M. Shahinpoor (eds) High Pressure Shock Compression of Solids II:
  Dynamic Fracture and Fragmentation}}}\ (\bibinfo  {publisher} {Springer, New
  York},\ \bibinfo {year} {1996})\ pp.\ \bibinfo {pages} {219--236}\BibitemShut
  {NoStop}%
\bibitem [{\citenamefont {Bakshi}(2006)}]{Bakshi2006}%
  \BibitemOpen
  \bibfield  {author} {\bibinfo {author} {\bibfnamefont {V.}~\bibnamefont
  {Bakshi}},\ }\href@noop {} {\emph {\bibinfo {title} {EUV Sources for
  Lithography}}},\ Press Monographs\ (\bibinfo  {publisher} {SPIE Press},\
  \bibinfo {year} {2006})\BibitemShut {NoStop}%
\bibitem [{\citenamefont {Mizoguchi}\ \emph {et~al.}(2015)\citenamefont
  {Mizoguchi}, \citenamefont {Nakarai}, \citenamefont {Abe}, \citenamefont
  {Nowak}, \citenamefont {Kawasuji}, \citenamefont {Tanaka}, \citenamefont
  {Watanabe}, \citenamefont {Hori}, \citenamefont {Kodama}, \citenamefont
  {Shiraishi}, \citenamefont {Yanagida}, \citenamefont {Soumagne},
  \citenamefont {Yamada}, \citenamefont {Yamazaki}, \citenamefont {Okazaki},\
  and\ \citenamefont {Saitou}}]{Mizoguchi_Nakari.2015}%
  \BibitemOpen
  \bibfield  {author} {\bibinfo {author} {\bibfnamefont {H.}~\bibnamefont
  {Mizoguchi}}, \bibinfo {author} {\bibfnamefont {H.}~\bibnamefont {Nakarai}},
  \bibinfo {author} {\bibfnamefont {T.}~\bibnamefont {Abe}}, \bibinfo {author}
  {\bibfnamefont {K.~M.}\ \bibnamefont {Nowak}}, \bibinfo {author}
  {\bibfnamefont {Y.}~\bibnamefont {Kawasuji}}, \bibinfo {author}
  {\bibfnamefont {H.}~\bibnamefont {Tanaka}}, \bibinfo {author} {\bibfnamefont
  {Y.}~\bibnamefont {Watanabe}}, \bibinfo {author} {\bibfnamefont
  {T.}~\bibnamefont {Hori}}, \bibinfo {author} {\bibfnamefont {T.}~\bibnamefont
  {Kodama}}, \bibinfo {author} {\bibfnamefont {Y.}~\bibnamefont {Shiraishi}},
  \bibinfo {author} {\bibfnamefont {T.}~\bibnamefont {Yanagida}}, \bibinfo
  {author} {\bibfnamefont {G.}~\bibnamefont {Soumagne}}, \bibinfo {author}
  {\bibfnamefont {T.}~\bibnamefont {Yamada}}, \bibinfo {author} {\bibfnamefont
  {T.}~\bibnamefont {Yamazaki}}, \bibinfo {author} {\bibfnamefont
  {S.}~\bibnamefont {Okazaki}}, \ and\ \bibinfo {author} {\bibfnamefont
  {T.}~\bibnamefont {Saitou}},\ }\bibfield  {title} {\enquote {\bibinfo {title}
  {Performance of one hundred watt {HVM LPP-EUV} source},}\ }\href {\doibase
  10.1117/12.2086347} {\bibfield  {journal} {\bibinfo  {journal} {Proc. SPIE}\
  }\textbf {\bibinfo {volume} {9422}},\ \bibinfo {pages} {94220C--94220C--13}
  (\bibinfo {year} {2015})}\BibitemShut {NoStop}%
\bibitem [{\citenamefont {Martynyuk}(1991)}]{Martynyuk1991}%
  \BibitemOpen
  \bibfield  {author} {\bibinfo {author} {\bibfnamefont {M.~M.}\ \bibnamefont
  {Martynyuk}},\ }\bibfield  {title} {\enquote {\bibinfo {title} {Generalized
  van der {W}aals equation of state for liquids and gases},}\ }\href@noop {}
  {\bibfield  {journal} {\bibinfo  {journal} {Zh. Fiz. Khim.}\ }\textbf
  {\bibinfo {volume} {65}},\ \bibinfo {pages} {1716--1717} (\bibinfo {year}
  {1991})}\BibitemShut {NoStop}%
\bibitem [{\citenamefont {Martynyuk}(1993)}]{Martynyuk1993}%
  \BibitemOpen
  \bibfield  {author} {\bibinfo {author} {\bibfnamefont {M.~M.}\ \bibnamefont
  {Martynyuk}},\ }\bibfield  {title} {\enquote {\bibinfo {title} {Transition of
  liquid metals into vapor in the process of pulse heating by current},}\
  }\href {\doibase 10.1007/BF00566045} {\bibfield  {journal} {\bibinfo
  {journal} {International Journal of Thermophysics}\ }\textbf {\bibinfo
  {volume} {14}},\ \bibinfo {pages} {457--470} (\bibinfo {year}
  {1993})}\BibitemShut {NoStop}%
\bibitem [{\citenamefont {Basko}(2018{\natexlab{b}})}]{Basko2018}%
  \BibitemOpen
  \bibfield  {author} {\bibinfo {author} {\bibfnamefont {M.~M.}\ \bibnamefont
  {Basko}},\ }\bibfield  {title} {\enquote {\bibinfo {title} {Generalized {van
  der Waals} equation of state for in-line use in hydrodynamic codes},}\ }\href
  {\doibase 10.20948/prepr-2018-112-e} {\bibfield  {journal} {\bibinfo
  {journal} {Keldysh Institute Preprints}\ }\textbf {\bibinfo {volume} {112}},\
  \bibinfo {pages} {28~p.} (\bibinfo {year} {2018}{\natexlab{b}})}\BibitemShut
  {NoStop}%
\bibitem [{\citenamefont {Haynes}(2015)}]{CRC2015}%
  \BibitemOpen
  \bibfield  {author} {\bibinfo {author} {\bibfnamefont {W.~M.}\ \bibnamefont
  {Haynes}},\ }\href@noop {} {\emph {\bibinfo {title} {CRC Handbook of
  Chemistry and Physics, 96th Edition}}},\ 100 Key Points\ (\bibinfo
  {publisher} {CRC Press},\ \bibinfo {year} {2015})\BibitemShut {NoStop}%
\bibitem [{\citenamefont {Landau}\ and\ \citenamefont
  {Lifshitz}(1996)}]{LL-SP96}%
  \BibitemOpen
  \bibfield  {author} {\bibinfo {author} {\bibfnamefont {L.~D.}\ \bibnamefont
  {Landau}}\ and\ \bibinfo {author} {\bibfnamefont {E.~M.}\ \bibnamefont
  {Lifshitz}},\ }\href@noop {} {\emph {\bibinfo {title} {{S}tatistical
  {P}hysics}}},\ \bibinfo {edition} {3rd}\ ed.\ (\bibinfo  {publisher}
  {Butterworth Heinemann},\ \bibinfo {year} {1996})\BibitemShut {NoStop}%
\bibitem [{\citenamefont {Landau}\ and\ \citenamefont
  {Lifshitz}(1987)}]{LL-H87}%
  \BibitemOpen
  \bibfield  {author} {\bibinfo {author} {\bibfnamefont {L.~D.}\ \bibnamefont
  {Landau}}\ and\ \bibinfo {author} {\bibfnamefont {E.~M.}\ \bibnamefont
  {Lifshitz}},\ }\href@noop {} {\emph {\bibinfo {title} {Fluid Mechanics}}},\
  Course of theoretical physics\ (\bibinfo  {publisher} {Pergamon Press},\
  \bibinfo {year} {1987})\BibitemShut {NoStop}%
\bibitem [{\citenamefont {Kanel}(2010)}]{Kanel2010}%
  \BibitemOpen
  \bibfield  {author} {\bibinfo {author} {\bibfnamefont {G.~I.}\ \bibnamefont
  {Kanel}},\ }\bibfield  {title} {\enquote {\bibinfo {title} {Spall fracture:
  methodological aspects, mechanisms and governing factors},}\ }\href {\doibase
  10.1007/s10704-009-9438-0} {\bibfield  {journal} {\bibinfo  {journal}
  {International Journal of Fracture}\ }\textbf {\bibinfo {volume} {163}},\
  \bibinfo {pages} {173--191} (\bibinfo {year} {2010})}\BibitemShut {NoStop}%
\bibitem [{\citenamefont {Imre}\ \emph {et~al.}(2008)\citenamefont {Imre},
  \citenamefont {Drozd-Rzoska}, \citenamefont {Kraska}, \citenamefont
  {Rzoska},\ and\ \citenamefont {Wojciechowski}}]{Imre_Drozd-Rzoska.2008}%
  \BibitemOpen
  \bibfield  {author} {\bibinfo {author} {\bibfnamefont {A.~R.}\ \bibnamefont
  {Imre}}, \bibinfo {author} {\bibfnamefont {A.}~\bibnamefont {Drozd-Rzoska}},
  \bibinfo {author} {\bibfnamefont {T.}~\bibnamefont {Kraska}}, \bibinfo
  {author} {\bibfnamefont {S.~J.}\ \bibnamefont {Rzoska}}, \ and\ \bibinfo
  {author} {\bibfnamefont {K.~W.}\ \bibnamefont {Wojciechowski}},\ }\bibfield
  {title} {\enquote {\bibinfo {title} {Spinodal strength of liquids, solids and
  glasses},}\ }\href {\doibase 10.1088/0953-8984/20/24/244104} {\bibfield
  {journal} {\bibinfo  {journal} {Journal of Physics: Condensed Matter}\
  }\textbf {\bibinfo {volume} {20}},\ \bibinfo {pages} {244104} (\bibinfo
  {year} {2008})}\BibitemShut {NoStop}%
\bibitem [{\citenamefont {Malyshev}\ \emph {et~al.}(2015)\citenamefont
  {Malyshev}, \citenamefont {Marin}, \citenamefont {Moiseeva}, \citenamefont
  {Gumerov},\ and\ \citenamefont {Akhatov}}]{Malyshev_Marin.2015}%
  \BibitemOpen
  \bibfield  {author} {\bibinfo {author} {\bibfnamefont {V.~L.}\ \bibnamefont
  {Malyshev}}, \bibinfo {author} {\bibfnamefont {D.~F.}\ \bibnamefont {Marin}},
  \bibinfo {author} {\bibfnamefont {E.~F.}\ \bibnamefont {Moiseeva}}, \bibinfo
  {author} {\bibfnamefont {N.~A.}\ \bibnamefont {Gumerov}}, \ and\ \bibinfo
  {author} {\bibfnamefont {I.~S.}\ \bibnamefont {Akhatov}},\ }\bibfield
  {title} {\enquote {\bibinfo {title} {Study of the tensile strength of a
  liquid by molecular dynamics methods},}\ }\href {\doibase
  10.1134/S0018151X15020145} {\bibfield  {journal} {\bibinfo  {journal} {High
  Temperature}\ }\textbf {\bibinfo {volume} {53}},\ \bibinfo {pages} {406--412}
  (\bibinfo {year} {2015})}\BibitemShut {NoStop}%
\bibitem [{\citenamefont {Blander}\ and\ \citenamefont
  {Katz}(1975)}]{Blander_Katz1975}%
  \BibitemOpen
  \bibfield  {author} {\bibinfo {author} {\bibfnamefont {M.}~\bibnamefont
  {Blander}}\ and\ \bibinfo {author} {\bibfnamefont {J.~L.}\ \bibnamefont
  {Katz}},\ }\bibfield  {title} {\enquote {\bibinfo {title} {{B}ubble
  nucleation in liquids},}\ }\href {\doibase 10.1002/aic.690210502} {\bibfield
  {journal} {\bibinfo  {journal} {AIChE Journal}\ }\textbf {\bibinfo {volume}
  {21}},\ \bibinfo {pages} {833--848} (\bibinfo {year} {1975})}\BibitemShut
  {NoStop}%
\bibitem [{\citenamefont {Skripov}\ and\ \citenamefont
  {Skripov}(1979)}]{Skripov_Skripov1979}%
  \BibitemOpen
  \bibfield  {author} {\bibinfo {author} {\bibfnamefont {V.~P.}\ \bibnamefont
  {Skripov}}\ and\ \bibinfo {author} {\bibfnamefont {A.~V.}\ \bibnamefont
  {Skripov}},\ }\bibfield  {title} {\enquote {\bibinfo {title} {Spinodal
  decomposition (phase transitions via unstable states)},}\ }\href {\doibase
  10.1070/PU1979v022n06ABEH005571} {\bibfield  {journal} {\bibinfo  {journal}
  {Sov. Phys. Usp.}\ }\textbf {\bibinfo {volume} {22}},\ \bibinfo {pages} {389}
  (\bibinfo {year} {1979})}\BibitemShut {NoStop}%
\bibitem [{\citenamefont {Skripov}(1974)}]{Skripov1974}%
  \BibitemOpen
  \bibfield  {author} {\bibinfo {author} {\bibfnamefont {V.~P.}\ \bibnamefont
  {Skripov}},\ }\href@noop {} {\emph {\bibinfo {title} {{M}etastable
  {L}iquids}}}\ (\bibinfo  {publisher} {Wiley, New York},\ \bibinfo {year}
  {1974})\BibitemShut {NoStop}%
\bibitem [{\citenamefont {Dekel}\ \emph {et~al.}(1998)\citenamefont {Dekel},
  \citenamefont {Eliezer}, \citenamefont {Henis}, \citenamefont {Moshe},
  \citenamefont {Ludmirsky},\ and\ \citenamefont
  {Goldberg}}]{Dekel_Eliezer.1998}%
  \BibitemOpen
  \bibfield  {author} {\bibinfo {author} {\bibfnamefont {E.}~\bibnamefont
  {Dekel}}, \bibinfo {author} {\bibfnamefont {S.}~\bibnamefont {Eliezer}},
  \bibinfo {author} {\bibfnamefont {Z.}~\bibnamefont {Henis}}, \bibinfo
  {author} {\bibfnamefont {E.}~\bibnamefont {Moshe}}, \bibinfo {author}
  {\bibfnamefont {A.}~\bibnamefont {Ludmirsky}}, \ and\ \bibinfo {author}
  {\bibfnamefont {I.~B.}\ \bibnamefont {Goldberg}},\ }\bibfield  {title}
  {\enquote {\bibinfo {title} {Spallation model for the high strain rates
  range},}\ }\href {\doibase 10.1063/1.368727} {\bibfield  {journal} {\bibinfo
  {journal} {Journal of Applied Physics}\ }\textbf {\bibinfo {volume} {84}},\
  \bibinfo {pages} {4851--4858} (\bibinfo {year} {1998})}\BibitemShut {NoStop}%
\bibitem [{\citenamefont {Faik}\ \emph {et~al.}(2012)\citenamefont {Faik},
  \citenamefont {Basko}, \citenamefont {Tauschwitz}, \citenamefont
  {Iosilevskiy},\ and\ \citenamefont {Maruhn}}]{Faik_Basko.2012}%
  \BibitemOpen
  \bibfield  {author} {\bibinfo {author} {\bibfnamefont {S.}~\bibnamefont
  {Faik}}, \bibinfo {author} {\bibfnamefont {M.~M.}\ \bibnamefont {Basko}},
  \bibinfo {author} {\bibfnamefont {A.}~\bibnamefont {Tauschwitz}}, \bibinfo
  {author} {\bibfnamefont {I.}~\bibnamefont {Iosilevskiy}}, \ and\ \bibinfo
  {author} {\bibfnamefont {J.~A.}\ \bibnamefont {Maruhn}},\ }\bibfield  {title}
  {\enquote {\bibinfo {title} {Dynamics of volumetrically heated matter passing
  through the liquid-vapor metastable states},}\ }\href {\doibase
  10.1016/j.hedp.2012.08.003} {\bibfield  {journal} {\bibinfo  {journal} {High
  Energy Density Phys.}\ }\textbf {\bibinfo {volume} {8}},\ \bibinfo {pages}
  {349--359} (\bibinfo {year} {2012})}\BibitemShut {NoStop}%
\bibitem [{\citenamefont {Basko}(2001)}]{Basko2001-DEIRA}%
  \BibitemOpen
  \bibfield  {author} {\bibinfo {author} {\bibfnamefont {M.~M.}\ \bibnamefont
  {Basko}},\ }\href {http://www.basko.net/mm/deira} {\enquote {\bibinfo {title}
  {{DEIRA}: {A} 1-{D} 3-{T} hydrodynamic code for simulating {ICF} targets
  driven by fast ion beams. {V}ersion 4},}\ } (\bibinfo {year} {2001}),\
  \bibinfo {note} {(accessed 27 September 2020; unpublished)}\BibitemShut
  {NoStop}%
\bibitem [{\citenamefont {Basko}(1990)}]{Basko1990}%
  \BibitemOpen
  \bibfield  {author} {\bibinfo {author} {\bibfnamefont {M.~M.}\ \bibnamefont
  {Basko}},\ }\bibfield  {title} {\enquote {\bibinfo {title} {Spark and volume
  ignition of {DT} and {D2} microspheres},}\ }\href {\doibase
  10.1088/0029-5515/30/12/001} {\bibfield  {journal} {\bibinfo  {journal}
  {Nuclear Fusion}\ }\textbf {\bibinfo {volume} {30}},\ \bibinfo {pages}
  {2443--2452} (\bibinfo {year} {1990})}\BibitemShut {NoStop}%
\bibitem [{\citenamefont {Fortov}, \citenamefont {Iakubov},\ and\ \citenamefont
  {Khrapak}(2006)}]{Fortov_Iakubov.2006}%
  \BibitemOpen
  \bibfield  {author} {\bibinfo {author} {\bibfnamefont {V.}~\bibnamefont
  {Fortov}}, \bibinfo {author} {\bibfnamefont {I.}~\bibnamefont {Iakubov}}, \
  and\ \bibinfo {author} {\bibfnamefont {A.}~\bibnamefont {Khrapak}},\
  }\href@noop {} {\emph {\bibinfo {title} {Physics of {S}trongly {C}oupled
  {P}lasma}}},\ \bibinfo {edition} {1st}\ ed.,\ International series of
  monographs on physics\ (\bibinfo  {publisher} {Clarendon Press, Oxford},\
  \bibinfo {year} {2006})\BibitemShut {NoStop}%
\bibitem [{\citenamefont {Thompson}(1971)}]{Thomson1971}%
  \BibitemOpen
  \bibfield  {author} {\bibinfo {author} {\bibfnamefont {P.~A.}\ \bibnamefont
  {Thompson}},\ }\bibfield  {title} {\enquote {\bibinfo {title} {A fundamental
  derivative in gasdynamics},}\ }\href {\doibase 10.1063/1.1693693} {\bibfield
  {journal} {\bibinfo  {journal} {Phys. Fluids (1958-1988)}\ }\textbf {\bibinfo
  {volume} {14}},\ \bibinfo {pages} {1843--1849} (\bibinfo {year}
  {1971})}\BibitemShut {NoStop}%
\bibitem [{\citenamefont {Mayer}\ and\ \citenamefont
  {Mayer}(2015)}]{Mayer_Mayer2015}%
  \BibitemOpen
  \bibfield  {author} {\bibinfo {author} {\bibfnamefont {A.~E.}\ \bibnamefont
  {Mayer}}\ and\ \bibinfo {author} {\bibfnamefont {P.~N.}\ \bibnamefont
  {Mayer}},\ }\bibfield  {title} {\enquote {\bibinfo {title} {Continuum model
  of tensile fracture of metal melts and its application to a problem of
  high-current electron irradiation of metals},}\ }\href {\doibase
  10.1063/1.4926861} {\bibfield  {journal} {\bibinfo  {journal} {Journal of
  Applied Physics}\ }\textbf {\bibinfo {volume} {118}},\ \bibinfo {pages}
  {035903} (\bibinfo {year} {2015})}\BibitemShut {NoStop}%
\bibitem [{\citenamefont {Luo}\ \emph {et~al.}(2009)\citenamefont {Luo},
  \citenamefont {An}, \citenamefont {Germann},\ and\ \citenamefont
  {Han}}]{Luo_An.2009}%
  \BibitemOpen
  \bibfield  {author} {\bibinfo {author} {\bibfnamefont {S.-N.}\ \bibnamefont
  {Luo}}, \bibinfo {author} {\bibfnamefont {Q.}~\bibnamefont {An}}, \bibinfo
  {author} {\bibfnamefont {T.~C.}\ \bibnamefont {Germann}}, \ and\ \bibinfo
  {author} {\bibfnamefont {L.-B.}\ \bibnamefont {Han}},\ }\bibfield  {title}
  {\enquote {\bibinfo {title} {Shock-induced spall in solid and liquid cu at
  extreme strain rates},}\ }\href {\doibase 10.1063/1.3158062} {\bibfield
  {journal} {\bibinfo  {journal} {Journal of Applied Physics}\ }\textbf
  {\bibinfo {volume} {106}},\ \bibinfo {pages} {013502} (\bibinfo {year}
  {2009})}\BibitemShut {NoStop}%
\bibitem [{\citenamefont {Zel'dovich}\ and\ \citenamefont
  {Raizer}(2012)}]{Zeldovich_Raizer2012}%
  \BibitemOpen
  \bibfield  {author} {\bibinfo {author} {\bibfnamefont {Y.~B.}\ \bibnamefont
  {Zel'dovich}}\ and\ \bibinfo {author} {\bibfnamefont {Y.~P.}\ \bibnamefont
  {Raizer}},\ }\href@noop {} {\emph {\bibinfo {title} {Physics of Shock Waves
  and High-Temperature Hydrodynamic Phenomena}}},\ Dover Books on Physics\
  (\bibinfo  {publisher} {Dover Publications},\ \bibinfo {year}
  {2012})\BibitemShut {NoStop}%
\end{thebibliography}%

\end{document}